\documentclass[12pt]{iopart}
\usepackage{bm}% bold math
\usepackage{iopams}
\usepackage{graphicx,epsfig}

\newcommand{\text}[1]{\mbox{\scriptsize{#1}}}

\begin{document}

\title{The Kovacs effect: a master equation analysis}
\author{A. Prados and J. J. Brey}
\address{F\'{\i}sica Te\'{o}rica, Universidad de Sevilla,
Apartado de Correos 1065, E-41080, Sevilla, Spain}
\eads{\mailto{prados@us.es},\mailto{brey@us.es}}

\begin{abstract}
The Kovacs or crossover effect is one of the peculiar behaviours exhibited by glasses and other complex, slowly relaxing  systems. Roughly it consists in the non-monotonic relaxation to its equilibrium value of a macroscopic property of a system evolving at constant temperature, when starting from a non-equilibrium state. Here, this effect is investigated for general systems whose dynamics is described by a master equation. To carry out a detailed analysis, the limit of small perturbations in which linear response theory applies is considered. It is shown that, under very general conditions, the observed experimental features of the Kovacs effect are recovered. The results are particularized for a very simple model, a two-level system with dynamical disorder. An explicit analytical expression for its non-monotonic relaxation function is obtained, showing a resonant-like behaviour when the dependence on the temperature is investigated.
\end{abstract}


\maketitle

\section{Introduction}
\label{s0}

Glassy behaviour has been quite an active field of investigation in recent years. A review of the main, experimentally observed features of real glasses, as well as of  several models showing dynamical properties analogous to them can be found in  \cite{Sch86,ANMMyM00}. In relaxation experiments, linear response functions usually exhibit  non-exponential decay. In particular, a Kohlrausch-Williams-Watts (KWW) law is often found. Moreover, a laboratory glass transition, in which the properties characterizing the macroscopic state of the system become frozen, is also observed. The transition is associated to a fast increase of the relaxation times upon decreasing the temperature. Along reheating, hysteresis effects show up, with the system going back to equilibrium following a curve which differs from the equilibrium one. In these reheating experiments, the curve overshoots equilibrium, and the difference between the actual value of the property of interest and its equilibrium value as a function of the temperature shows a non-monotonic behaviour. This is a typical example of a memory effect: the behaviour of the system depends on its entire thermal history, and not only on the instantaneous, initial value of the property under study.

\begin{figure}
\begin{center}
\epsfig{file=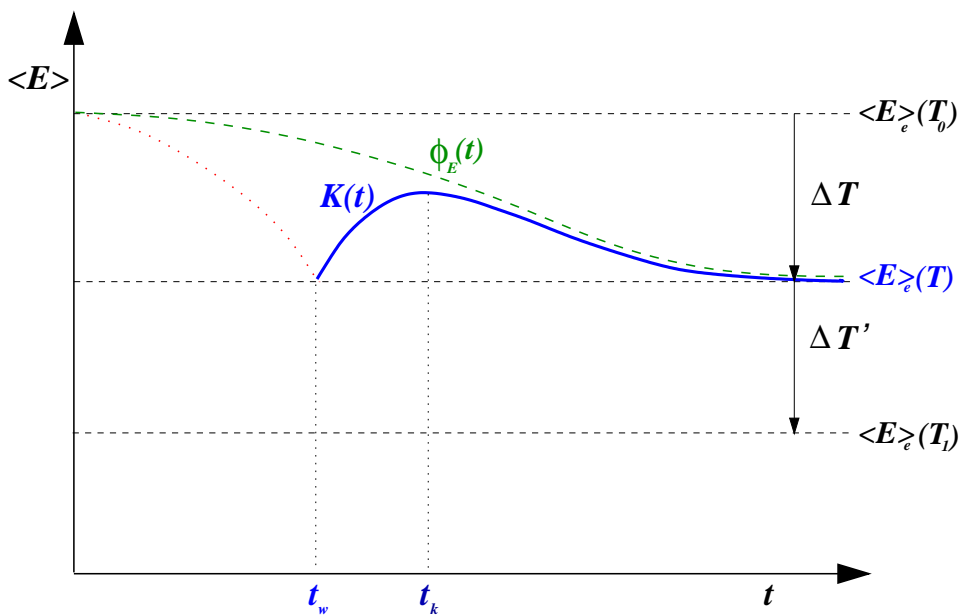, width=5in}
\caption[]{Schematic representation of the Kovacs experiment described in the text. The dashed green curve $\phi_E(t)$ represents the direct relaxation from $T_0$ to $T$. The dotted red curve stands for the part of the relaxation from $T_0$ to $T_1$, which is interrupted by the second temperature jump, changing abruptly the temperature from $T_1$ to $T$ at $t=t_w$. After this second jump, the system follows the solid blue curve $K(t)$, which reaches a maximum for $t=t_k$ and, afterwards, approaches $\phi_E(t)$ for very long times.\label{fig1}}
\end{center}
\end{figure}

One of the first and simplest experiments revealing a memory effect was designed by Kovacs \cite{Ko63,Ko79}. Polyvinyl acetate was equilibrated at a high temperature $T_0$, and the ``direct'', monotonic, relaxation function $\phi (t)$ of the volume to a low temperature $T<T_0$ was measured. Then, after equilibrating the system again at $T_0$, it was rapidly quenched, this time to a lower temperature $T_1<T$, at which it isothermally relaxed for a time $t_w$. This time $t_w$ was not enough for the system to reach its equilibrium state at temperature $T_1$. Then, the temperature was abruptly increased to $T$. The time $t_{w}$ was chosen  such that the volume at $t_w$ was equal to the equilibrium value at $T$. The observed behaviour of the system for $t>t_w$ turned out to be quite peculiar. The volume did not remain constant, but it first increased, passed through a maximum at a certain time $t_k>t_w$,  relaxing afterwards to the equilibrium value at $T$. The results of the Kovacs experiment are qualitatively sketched in figure \ref{fig1}, although the property considered there is the energy $E$, instead of the volume $V$, because the former is the  quantity that will be studied in this paper. The non-monotonic behaviour for $t>t_w$ is represented by the curve $K(t)$, while  $\phi_E(t)$ describes the direct relaxation of the energy from $T_0$ to $T$. In the  Kovacs experiment,  the pressure $P$ was kept fixed along all the processes. Then the observed behaviour  means that the knowledge of the state variables, $P$, $V$, and $T$, does not fully characterize the macroscopic state of the system, since the subsequent evolution is different for systems with the same values of the state variables but with different thermal history. This is the reason why this experiment is said to show a memory effect. Besides, this non-monotonic behaviour, sometimes called the Kovacs ``hump'', displays some characteristic features. For fixed initial and final temperatures $T_0$ and $T$, the magnitude of the maximum increases as the intermediate temperature $T_1$  decreases, i.e. the maximum is higher the larger  the second temperature jump $T-T_1$. Besides, the position of the maximum moves to the left, in the sense that $t_k-t_w$ is a decreasing function of $T-T_1$. Finally, for very long times, when the Kovacs hump function $K(t)$ decreases towards the equilibrium value at $T$, it approaches the direct relaxation curve $\phi_E$.

The Kovacs effect has been investigated in many systems showing slow glassy dynamics. Kovacs himself presented a phenomenological description in \cite{Ko79}. Later on, it has been analyzed both analytically and numerically in several models. This includes a molecular dynamics analysis of a realistic model of ortho-terphenyl \cite{MyS04}, simulations of a 3d spin-glass \cite{ByB02}, of classical and quantum p-spin models \cite{CLyL04}, and of kinetically constrained models \cite{Bu03,AyS04}. Also, analytical investigations have been carried out in the context of simple models. Examples are the Glauber 1d Ising model \cite{Br78}, the 2d XY model \cite{ByH02}, domain growth and trap models \cite{BBDyG03}, the parking lot model for granular compaction \cite{TyV04}, the HOSS models for a fragile glass \cite{ALyN06}, and a distribution of two-level systems \cite{AAyN08}. All the above systems exhibit a behaviour resembling the experimental Kovacs ``hump'', with a maximum whose height increases with the temperature jump $T-T_1$. The time $t_k$ for the maximum usually decreases with the jump $T-T_1$, although in some models $t_k-t_w$ appears to be almost independent of the jump and, also, the tendency towards the direct quench curve is not clear \cite{TyV04,AAyN08}.

In this work, we would like to improve the current understanding of the Kovacs effect by trying to identify the origin of its generality. In particular, one of the main aims of this work is to prove that it can be considered as a general feature of systems whose dynamics is described by a master equation, i.e. homogenous Markov systems. Some of the models mentioned above belong to this kind of systems \cite{ByB02,Bu03,AyS04,Br78,ALyN06,AAyN08}.
Besides, we will assume that the transition rates verify detailed balance, although the results can be also valid in some cases in which this condition is not fulfilled, as discussed in the last section of the paper. In order to carry out a detailed analytical analysis, the temperature jumps will be considered small enough, so that linear response theory can be applied. It will be shown that, in this limit, the Kovacs effect as well as its main features are still present, clearly indicating that it is not a non-linear effect.

The plan of the paper is as follows. Section \ref{s1} is devoted to a general discussion of the linear response to a temperature jump of a system whose dynamics is described by a master equation. The obtained results are applied to the Kovacs experiment in section \ref{s2}, and a general expression for the Kovacs ``hump'' function $K(t)$ is derived. Some general features of this function are discussed in section \ref{s2a}. In section \ref{s3}, the two level system with dynamical disorder is introduced, and some of its properties are shortly reviewed. The Kovacs effect in this system is investigated in section \ref{s3a}. The simplicity of the model allows us  to obtain an explicit analytic expression for $K(t)$, which is analyzed in the light of the general results obtained previously. Section \ref{s4} contains a summary of some of the main results and conclusions of the present work. Finally, some calculations are presented in the two appendices.

\section{Linear response to a small temperature jump}
\label{s1}

Consider a system whose dynamics is described by a homogenous Markov process. The conditional probability $T_{t-t^\prime}(i|j)$ of finding the system in state $i$ at time $t$, given that it was in state $j$ at time $t^\prime<t$, obeys the master equation
\begin{equation}
\label{1.1}
\frac{\partial}{\partial t} T_{t-t^\prime}(i|j)=\sum_k \left[ W_{ik} T_{t-t^\prime}(k|j)-W_{ki} T_{t-t^\prime}(i|j) \right] \, .
\end{equation}
Here  $W_{ik}$ is the (time-independent) transition rate from state $k$ to state $i$. The above equation has to be solved with the initial condition $T_{0}(i|j) = \delta_{ij}$. The probability ${\cal P}_{i}(t)$ that the system be in state $i$ at time $t$ is obtained from the conditional probability and some initial condition ${\cal P}_{i}(0)$ through
\begin{equation}
\label{1.1a}
{\cal P}_{i}(t) = \sum_{j} T_{t}(i|j) {\cal P}_{j}(0).
\end{equation}
The set of both $T_{t-t^\prime}(i|j)$ and ${\cal P}_i(t)$ completely determines the Markov process. From (\ref{1.1}) and (\ref{1.1a}) it follows that ${\cal P}_{i}(t)$ also obeys the master equation,
\begin{equation}\label{1.2}
    \frac{\partial }{ \partial t}{\cal P}_i(t)=\sum_k \left[ W_{ik} {\cal P}_k(t)-W_{ki} {\cal P}_i(t) \right].
\end{equation}
It is convenient to introduce the matrix   $\mathbb{W}$ defined by \cite{vk92}
\begin{equation}\label{1.3}
    \mathbb{W}_{ij} \equiv W_{ij}- \delta_{ij} \sum_k W_{kj} \, ,
\end{equation}
so that  (\ref{1.2})  can be written in matrix notation as
\begin{equation}\label{1.4}
    \frac{ \partial}{\partial t} {\cal P} (t)=\mathbb{W} {\cal P}(t) \, ,
\end{equation}
where ${\cal P} (t)$ is the column matrix with elements ${\cal P}_i (t)$. The formal solution of this equation is
\begin{equation}\label{1.5}
    {\cal P}(t)=e^{t \mathbb{W}} {\cal P}(0) \,
\end{equation}
or, in component notation,
\begin{equation}\label{1.6}
    {\cal P}_i(t)=\sum_j \left( e^{t \mathbb{W}} \right)_{ij} {\cal P}_j(0).
\end{equation}
Equivalently, the solution of (\ref{1.1}) is
\begin{equation}\label{1.7}
    T_{t-t^{\prime}}(i|j)=\left( e^{(t-t^{\prime}) \mathbb{W}} \right)_{ij} \, .
\end{equation}

The transition rates often depend on a set of macroscopic parameters $\mathbf{X}=\left\{X_1,X_2,\ldots, X_{r}\right\}$, such as the temperature, pressure, density, external fields, and so on. In some cases,  the values of the above parameters can be externally controlled. Along this paper, attention will be restricted to this sort of situations. Suppose that a system is evolving being
 $\mathbf{X^{(0)}} \equiv\left\{ X_1^{(0)},X_2^{(0)},\ldots,X_{r}^{(0)} \right\}$ the values of the macroscopic parameters
 and $\mathbb{W}^{(0)}\equiv \mathbb{W}(\mathbf{X^{(0)}})$ the corresponding evolution matrix. The
 probability distribution $\mathcal{P}^{(0)}(t)$ will obey the master equation,
\begin{equation}\label{1.10}
    \frac{ \partial \mathcal{P}^{(0)}(t)}{\partial t}=\mathbb{W}^{(0)}\mathcal{P}^{(0)}(t)\, .
\end{equation}
Next, consider the evolution of the same system, but with parameters given by  $\mathbf{X}= \mathbf{X}^{(0)}+ \Delta \mathbf{X}$. The new transition matrix can be decomposed as
\begin{equation}\label{1.11}
    \mathbb{W}  \equiv \mathbb{W}(\mathbf{X})=\mathbb{W}^{(0)}+\mathbb{W}^{\prime} \, ,
\end{equation}
where,
\begin{equation}\label{1.12}
    \mathbb{W}^{\prime}=
    \mathbb{W}(\mathbf{X^{(0)}}+\Delta\mathbf{X})-\mathbb{W}(\mathbf{X^{(0)}}) \, .
\end{equation}
The new probability distribution $\mathcal{P}(t)$ will obey (\ref{1.4}), and it can be decomposed in the form
\begin{equation}\label{1.15}
    \mathcal{P}(t)=\mathcal{P}^{(0)}(t)+\mathcal{P}^{\prime}(t) \, .
\end{equation}
Obviously, both $\mathbb{W}^{\prime}$ and $\mathcal{P}^{ \prime}$ vanish for $\Delta\mathbf{X}=0$. Substitution of
 (\ref{1.11}) and (\ref{1.15}) into  (\ref{1.4}) and use of  (\ref{1.10}) yields
\begin{equation}\label{1.14}
\frac{\partial}{\partial t} \mathcal{P}^{\prime}(t)=\mathbb{W}^{(0)}\mathcal{P}^{\prime} (t)+\mathbb{W}^{\prime} \mathcal{P}^{(0)}(t)+ \mathbb{W}^{\prime} \mathcal{P}^{\prime}(t).
\end{equation}
In the linear response approximation, the limit $ \Delta \mathbf{X} \rightarrow 0$ is considered, and it is assumed that the last term on the right hand side of  (\ref{1.14}) can be neglected, since it is at least of second order in the deviations $ \Delta \mathbf{X}$.

In the following, the above results will be particularized for an idealized kind of experiments. Starting from a given initial condition $\mathcal{P}(0)$, the system evolves for a time $t_{w}$ with values $\mathbf{X}^{(0)}$ of the macroscopic parameters. Then, at $t=t_w$ the parameters are instantaneously  changed to $\mathbf{X} = \mathbf{X}^{(0)} + \Delta \mathbf{X}$, being $\Delta \mathbf{X}$ very small. The probability distribution of the system for $t \geq t_{w}$ is given by (\ref{1.15}) and (\ref{1.14}). Integration of the latter after neglecting the nonlinear term  leads to
\begin{equation}\label{1.21}
    \mathcal{P}^{\prime}(t)=\int_{t_w}^t \rmd t^\prime e^{(t-t^{\prime}) \mathbb{W}^{(0)}} \mathbb{W}^{\prime} \mathcal{P}^{(0)}(t^\prime) \, ,
\end{equation}
valid for $t\geq t_w $, while $\mathcal{P}^{\prime}(t)=0$ for $t < t_{w}$. In component notation,
\begin{equation}\label{1.22}
    \mathcal{P}^{\prime}_i(t)=\int_{t_w}^t \rmd t^\prime \sum_j \sum_k T_{t-t^\prime}^{(0)}(i|j) \mathbb{W}^{\prime}_{jk} \mathcal{P}_k^{(0)}(t^\prime) \, ,
\end{equation}
$t\geq t_w \,$, with
\begin{equation}
\label{1.22a}
T_{t-t^\prime}^{(0)}(i|j)= \left( e^{ (t-t^{\prime}) \mathbb{W}^{(0)}} \right)_{ij}.
\end{equation}
The average value of property $A$ at time $t\geq t_w$, will be denoted by
\begin{equation}\label{1.23a}
    \langle A(t|t_w)\rangle\equiv\sum_i A_i \mathcal{P}_i(t),
\end{equation}
where $A_{i}$ is the value of the property $A$ in state $i$. From  (\ref{1.15}) and (\ref{1.22}) it follows that,
again for $t \geq t_{w}$,
\begin{equation}\label{1.23}
    \langle A(t|t_w)\rangle=\langle A(t)\rangle^{(0)}+\int_{t_w}^t \rmd t^\prime R_{A}(t,t^\prime) \, ,
\end{equation}
with
\begin{equation}\label{1.24}
    \langle A(t)\rangle^{(0)}=\sum_i A_i \mathcal{P}_i^{(0)}(t)
\end{equation}
and
\begin{equation}\label{1.25}
    R_{A}(t,t^\prime)=\sum_i \sum_j \sum_k A_i T_{t-t^\prime}^{(0)}(i|j) \mathbb{W}_{jk}^{\prime} \mathcal{P}_k^{(0)}(t^\prime).
\end{equation}
Equation (\ref{1.23}) implies that
\begin{equation}\label{1.26}
    \frac{\partial}{\partial t_w}\left[ \langle A(t|t_w)\rangle-\langle A(t)\rangle^{(0)}\right]= -R_{A}(t,t_w) \, .
\end{equation}
The above relations indicate that $R_{A}(t,t_w)$ can be interpreted as the linear response function in the property $A$ of the system to a modification of the external parameters $\mathbf{X}$. It is worth stressing that the system has not been assumed to be at equilibrium at time $t=t_{w}$, when the perturbation is switched on. Suppose now that this is the case, and that the equilibrium distribution of the system has the canonical form
\begin{equation}
\label{1.26a}
\mathcal{P}_{e,i}(T) \equiv \frac{e^{-\beta E_{i}}}{Z(T)},
\end{equation}
where $\beta \equiv (k_{B}T)^{-1}$, $k_{B}$ being the Boltzmann constant and $T$ the temperature. Moreover, $E_{i}$ is the energy of the system in state $i$ and $Z(T)$ the partition function
\begin{equation}
\label{1.26b}
Z(T) \equiv \sum_{i} e^{-\beta e_{i}}.
\end{equation}
 In addition, let us suppose that the only external parameter that is changed at $t =t_{w}$ is the temperature, which is modified from $T^{(0)}$ to $T^{(0)}+ \Delta T$ or, equivalently, $\beta^{(0)} \equiv (k_{B}T^{(0)})^{-1}$ is changed into $\beta^{(0)}+\Delta \beta$ with $\Delta \beta = -(k_{B}T^{(0)2})^{-1} \Delta T$. Under these circumstances, the response function is given by
\begin{equation}\label{1.27}
    R_{e,A}(t,t^\prime)=R_{e,A}(t-t^\prime)= \Delta\beta \frac{\partial}{\partial t} \langle A(t) E(t^\prime) \rangle_e^{(0)} \, ,
\end{equation}
$t\geq t^\prime\geq t_w$. Here $\langle A(t) B(t^\prime) \rangle_e^{(0)}$ denotes the equilibrium correlation function at temperature $T^{(0)}$,
\begin{equation}
\label{1.27a}
\langle A(t) B(t^\prime) \rangle_e^{(0)} \equiv \sum_{i} \sum_{j} A_{i}B_{j} T_{t-t^{\prime}}^{(0)} (i|j) \mathcal{P}_{e,j} (T^{(0)}).
\end{equation}
A proof of  (\ref{1.27}) is sketched in \ref{apa}.

\section{The Kovacs experiment}
\label{s2}

Let us  analyze the experiment proposed by Kovacs \cite{Ko63,Ko79} and  qualitatively depicted in figure \ref{fig1}. To characterize the non-monotonic behaviour of the average energy, let us decompose it for $t \geq t_{w}$ in the form
\begin{equation}\label{2.1}
    \langle E(t|t_{w})\rangle=\langle E\rangle_e(T)+K(t) \, ,
\end{equation}
where the Kovacs ``hump'' $K(t)$ satisfies
\begin{equation}\label{2.2}
    K(t) \geq 0 \quad \forall t \, , \quad \lim_{t\rightarrow t_{w}}K(t)=\lim_{t\rightarrow\infty} K(t)=0 \, .
\end{equation}
The specific shape of the hump depends both on the final and intermediate temperatures of the system, $T$ and $T_{1}$, respectively. It exhibits a maximum at a certain time $t_k$. For fixed $T$, the height of the maximum increases with the temperature jump $T-T_1$, while its position relative to $t_{w}$, $t_k-t_w$, decreases. For long enough times, $\langle E(t|t_{w})\rangle $ approaches the curve $\phi_E(t)$, corresponding to the relaxation at temperature $T$ starting from the equilibrium state at temperature $T_0$.

To render the problem tractable by analytical methods, the temperature jumps will be assumed to be small, i.e. $|\Delta T| \ll 1$ and $|\Delta T^{\prime}| \ll 1$, where  $\Delta T = T-T_{0}$ and $\Delta T^{\prime}=T_{1}-T$.
Then, the results derived in the previous section can be used. For the time interval $0<t<t_{w}$, the system relaxes at temperature $T_{1}=T+\Delta T^{\prime}$ starting from an equilibrium situation at $T^{(0)}$, and direct application of (\ref{1.23}) and (\ref{1.27}) gives
\begin{equation}\label{2.3}
    \langle E(t)\rangle=\langle E\rangle_e(T-\Delta T)-(\Delta\beta+\Delta\beta^\prime)\int_0^t \rmd t^\prime\, \frac{\partial}{\partial t^\prime} \langle E(t)E(t^\prime) \rangle_e \, ,
\end{equation}
$0<t<t_{w}$, with
\begin{equation}\label{2.6}
    \Delta\beta=- \frac{\Delta T}{k_B T^2}>0 \, , \quad \Delta\beta^\prime=- \frac{\Delta T^\prime}{k_B T^2}>0,
\end{equation}
In (\ref{2.3}), the equilibrium correlation function $\langle E(t)E(t^\prime)\rangle_{e}$ must be evaluated at temperature $T$, since calculating it at temperature $T_{0}$ would introduce nonlinear  corrections in the temperature jumps, which are neglected in the linear response theory we are using. In \ref{apb} it is shown that, for systems in which detailed balance holds, it is
\begin{equation}\label{2.7}
    \langle E(t)E(t^\prime)\rangle_e=\langle E\rangle_e^2+{\sum_\alpha}^\prime (\Delta E^{(\alpha)})^2 e^{-(t-t^{\prime})\lambda^{(\alpha)}} \, ,
\end{equation}
where $\alpha$ is the index for the eigenvectors of the matrix $\mathbb{W}$, $- \lambda ^{(\alpha)}$ is the eigenvalue corresponding to $\alpha$, and
\begin{equation}
\label{2.7a}
\Delta E^{(\alpha)} \equiv \sum_{i} E_{i} \psi_{i}^{(\alpha)}.
\end{equation}
Here $\psi_{i}^{(\alpha)}$ is the $i$-component of the eigenvector $\alpha$. The prime to the right of the summation symbol in  (\ref{2.7}) indicates that the equilibrium eigenvalue is excluded. Thus, the equilibrium time autocorrelation function of the energy is a monotonic decreasing function of the time interval $t-t^{\prime}$, since it is a linear combination of decaying exponentials with all its coefficients being positive. Particularization of
 (\ref{2.7}) for $t=t^\prime$ provides an interpretation for $\Delta E^{(\alpha)}$,
\begin{equation}\label{2.8}
    \sigma_E^2\equiv \langle E^2\rangle_e-\langle E\rangle_e^2 ={\sum_\alpha}^\prime (\Delta E^{(\alpha)})^2 \, ,
\end{equation}
i.e. $(\Delta E^{(\alpha)})^2$ is the contribution of the $\alpha$-th mode to the equilibrium energy dispersion. Equation (\ref{2.7}) can be expressed as
\begin{equation}\label{2.10}
    \langle E(t)E(t^\prime)\rangle_e-\langle E\rangle_e^2=\sigma_E^2 \phi_E(t-t^\prime) \, .
\end{equation}
with
\begin{equation}\label{2.11}
    \phi_E(t)={\sum_\alpha}^\prime b^{(\alpha)} e^{-\lambda^{(\alpha)} t} \,
\end{equation}
and
\begin{equation}\label{2.10b}
    b^{(\alpha)}=\frac{(\Delta E^{(\alpha)})^2 }{{\sum_{\alpha^\prime}}^\prime
    (\Delta E^{(\alpha^\prime)})^2 } \, .
\end{equation}
The weights $b^{(\alpha)}$ have the properties
\begin{equation}\label{2.10c}
    b^{(\alpha)}>0,   \quad {\sum_\alpha}^\prime b^{(\alpha)}=1 \, .
\end{equation}
This relaxation function $\phi_{E}(t)$ is ``normalized'', in the sense that $\phi_E(0)=1$,  and it verifies that $\phi_E(\infty)=0$ for all temperatures. Returning to  (\ref{2.3}), and using  (\ref{2.8}) and (\ref{2.10}),
\begin{eqnarray}\label{2.12}
\langle E(t) \rangle &=& \langle E\rangle_e(T)+ \sigma_E^2 \Delta\beta-(\Delta\beta+\Delta\beta^\prime) \sigma_E^2 \left[ 1-\phi_E(t) \right]  \nonumber \\
   &=& \langle E \rangle_e (T+ \Delta T^{\prime})+(\Delta\beta+\Delta\beta^\prime) \sigma_E^2 \phi_E(t)  \, .
\end{eqnarray}
This expression shows the monotonic relaxation of the average energy towards its equilibrium value at temperature $T_{1}=T+ \Delta T^{\prime}$. Now, the condition fixing the time $t_w$ at which the temperature is increased is employed,
\begin{equation}\label{2.13}
    \langle E(t_w)\rangle=\langle E\rangle_e(T)
\end{equation}
i.e., using  (\ref{2.12}),
\begin{equation}
\label{2.13a}     \phi_E(t_w)=\frac{\Delta\beta^\prime}{\Delta\beta+\Delta\beta^\prime}=\frac{|\Delta T^\prime|}{|\Delta T|+|\Delta T^\prime|}\equiv x \,.
\end{equation}
The last equality defines the variable $x$ taking values in the interval $0 \leq x \leq 1$. Accordingly with the Kovacs experiment described above, at $t=t_{w}$ the temperature is instantaneously raised from $T_{1}$ to $T=T_{1}- \Delta T^{\prime}$. To write the expression of the average energy for $t>t_{w}$,(\ref{1.23}) will be used again, but in this case the reference evolution is provided by  (\ref{2.12}), which describes the energy relaxation at temperature $T_{1}= T+ \Delta T^{\prime}$. Consequently, the response function $R_{E}(t,t^{\prime})$ to be used corresponds, in principle, to a non-equilibrium situation. However, since $R_{E}$ is proportional to $ \mathbb{W}^{\prime}$, it follows that, in the linear approximation, the distribution function in  (\ref{1.25}) can be replaced by the equilibrium distribution at temperature $T$. That means that the equilibrium response function has  to be considered again. In this way, it is easily obtained that
\begin{eqnarray}
    \langle E(t|t_{w})\rangle=\langle E \rangle_e (T+\Delta T^\prime)&+(\Delta\beta+\Delta\beta^\prime) \sigma_E^2 \phi_E(t) \nonumber \\
    &+\Delta\beta^\prime \int_{t_w}^t \rmd t^\prime \frac{\partial}{\partial t^\prime} \langle E(t)E(t^\prime) \rangle_e \, ,
\end{eqnarray}
and, taking into account again (\ref{2.10}),
\begin{equation}\label{2.14a}
    \langle E(t|t_{w}) \rangle=\langle E\rangle_e(T)+(\Delta\beta+\Delta\beta^\prime)\sigma_E^2 \phi_E(t)-\Delta\beta^\prime \sigma_E^2 \phi_E(t-t_w) \, .
\end{equation}
This equation is similar to the one obtained by Kovacs by means of a phenomenological theory \cite{Ko79}. The  difference is that, in the phenomenological theory, the arguments of the response functions are reduced times, in the spirit of Narayanaswami-Moynihan-Tool theory of supercooled liquids \cite{Na71,BEMyM76,To46}, instead of actual times as in  (\ref{2.14a}).

To characterize the evolution of the system for $t>t_w$, i.e. after the second temperature jump, it is convenient to use a dimensionless Kovacs hump (compare with (\ref{2.1})) by
\begin{equation}\label{2.14}
    K^{*}(t)=\frac{\langle E(t|t_{w})\rangle-\langle E\rangle_e(T)}{\sigma_E^2 \Delta\beta} \simeq \frac{\langle E(t|t_w)\rangle-\langle E\rangle_e(T)}{\langle E\rangle_e(T-\Delta T)-\langle E\rangle_e(T)} \, .
\end{equation}
This is the function which is analyzed in Kovacs' experiments \cite{Ko63,Ko79}. Although it is not explicitly shown in our notation, $K^*(t)$ is actually a function of both $t$ and the ``waiting'' time $t_w$. Therefore, $K^*(t_w,t)$ will also be used for the Kovacs hump when the dependence on $t_w$ should need to be emphasized.  Substitution of (\ref{2.14a}) into (\ref{2.14}) gives
\begin{equation}\label{2.15}
    K^{*}(t)=\frac{\phi_E(t)-x\phi_E(t-t_w)}{1-x} \, .
\end{equation}
where $x=\phi_E(t_w)$ is the function of the temperature jumps defined by (\ref{2.13a}).

In the next section, some general properties of this function will be derived. They follow from the form of the normalized energy autocorrelation function $\phi_E(t)$, as given by  (\ref{2.11}). For the analysis, it is useful to introduce the functions
\begin{eqnarray}\label{2.15b}
  D_1(t) & \equiv& -\frac{\rmd}{\rmd t}\ln\phi_E(t)=-\frac{\phi_E^\prime(t)}{\phi_E(t)} \ , \label{2.15b1} \\
  D_2(t) & \equiv & -\frac{\rmd}{\rmd t}\ln[-\phi^\prime_E(t)]=-\frac{\phi_E^{\prime\prime}(t)}{\phi_E^{\prime}(t)} \label{2.15b2} \, .
\end{eqnarray}
In the above expressions the prime is used to indicate derivative with respect to time. Both $D_1(t)$ and $D_2(t)$ are strictly positive decreasing functions of $t$, as a consequence of $\phi_E(t)$ being a linear combination of exponentials with positive coefficients. This implies  that $D_1(t)$ must tend to a well defined limit for $t\rightarrow\infty$,
\begin{equation}\label{2.15c}
    \lim_{t \rightarrow\infty} D_1(t)=D_1^{(\infty)}\geq 0 \, .
\end{equation}
If $D_1^{(\infty)}\neq 0$, the relaxation is basically exponential for very long times. This is compatible with $\phi_E(t)$ exhibiting  non-exponential behaviour in most of its relaxation, since the  values of $\phi_E(t)$ for which the exponential decay is observed can be very small, say $\phi_E(t)\leq 0.01$. This is the case, for instance, of the energy relaxation at low temperatures in the one dimensional Ising model with Glauber dynamics \cite{ByP93,ByP96}. On the other hand, if  $D_1^{(\infty)}=0$, the relaxation is non-exponential even for very large times. The above comments follow directly from the definition of $D_{1}(t)$ in  (\ref{2.15b1}).

\section{Properties of the relative response function $K^{*}(t)$}
\label{s2a}

\subsection{It is bounded between $0$ and  $\phi_{E}(t)$ for all $t\geq t_w$}
To begin with, it will be shown that for those times $t_{1}$ such that $K^{*}(t_{1})=0$, it is $K^{* \prime}(t_{1}) \geq 0$, i.e. $K^{*}(t)$ never crosses the time axis with a negative slope.  From  (\ref{2.15}),
\begin{equation}\label{2.16}
    K^{*\prime}(t)=\frac{ \phi_E^\prime(t)-x\phi_E^\prime(t-t_w)}{1-x} \, .
\end{equation}
If $K^{*}(t_1)=0$,  equation (\ref{2.15}) implies that $x=\phi_E(t_1)/\phi_E(t_1-t_w)$, and substitution of this into (\ref{2.16}) leads to
\begin{equation}\label{2.17}
    K^{* \prime}(t_1)=\frac{\phi_E(t_1)\left[D_1(t_1-t_{w})-D_1(t_1)\right]}{1-x}.
\end{equation}
Since the function $D_1(t)$ is a monotonically decreasing function of $t$, the term inside the square brackets is positive and, therefore,
\begin{equation}\label{2.19}
    K^{*\prime} (t_1) \geq 0 \, ,
\end{equation}
as indicated above. Moreover, the equality sign only holds if $\phi_E(t_1)=0$, i.e. for $t_1\rightarrow\infty$. As $K^{*}(t)$ vanishes by definition at $t=t_{w}$, it follows that $K^{*}(t)>0$ for all $t>t_w$,  tending to zero in the limit $t \rightarrow \infty$. In addition, $K^{*}(t)$ has an upper bound that follows directly  from  (\ref{2.15}), by taking into account that  $\phi_E(t)\leq\phi_E(t-t_w)$,
\begin{equation}\label{2.20}
    K^{*}(t)\leq \phi_E(t) \, .
\end{equation}
In the experiments \cite{Ko63,Ko79}, it has been observed that $K^*(t)\leq \phi_E(t-t_w)$. As $\phi_E(t)\leq\phi_E(t-t_w)$, we have derived here a more restrictive inequality, perhaps as a consequence of
having restricted ourselves to the linear response regime.

\subsection{There is only one maximum of $K^{*}(t)$}

As $K^{*}(t)$ is a regular positive function of $t$, vanishing both at $t=t_w$ and for $t\rightarrow\infty$, it must exhibit at least a maximum for $t>t_w$. It will be proved that actually there is only one. Suppose that a stationary point occurs at $t=t_{k}$, so that
\begin{equation}\label{2.21}
    K^{* \prime}(t_k)=\frac{\phi_E^\prime(t_k)-x\phi_E^\prime(t_k-t_w)}{1-x}=0 \, ,
\end{equation}
where  (\ref{2.16}) has been used. Equation (\ref{2.21}) defines $t_k$ as a function of $t_w$, i.e. of the ratio of the temperature jumps $|\Delta T|/|\Delta T^\prime|$. The time $t_k$ will correspond to a maximum or a minimum of $K^*(t)$ depending on the sign of the second derivative at $t_k$,
\begin{equation}\label{2.22a}
    K^{*\prime\prime}(t_k)=\frac{1}{1-x}\left[ \phi_E^{\prime\prime}(t_k)-x\phi_E^{\prime\prime}(t_k-t_w)\right] \, .
\end{equation}
 (\ref{2.21}) gives that $x=\phi_E^\prime(t_k)/\phi_E^{\prime}(t_k-t_w)$, which leads to
\begin{equation}\label{2.22}
    K^{*\prime\prime}(t_k)=\frac{\phi_E^\prime(t_k)}{1-x}\left[D_2(t_k-t_{w})-D_2(t_k)\right] \, ,
\end{equation}
where we have introduced the function $D_2(t)$ defined in  (\ref{2.15b2}). The term in brackets is again strictly positive, since $D_2(t)$ is also a positive monotonically decreasing function of $t$. Therefore, taking into account that $\phi_{E}$ is a monotonic decreasing function of time, it follows that
\begin{equation}\label{2.22b}
    K^{*\prime\prime}(t_k)\leq 0 \, ,
\end{equation}
and $t_k$ must correspond to a maximum. Moreover $K^{*\prime\prime}(t_k)=0$ is only possible if $\phi_E^\prime(t_k)=0$, i.e. for $t_k\rightarrow\infty$, when $K^*(t)$ approaches zero with horizontal tangent. It is also clear that there can only be one maximum of $K^*(t)$ since, on account of continuity, between two maximums there should be at least a minimum.

\subsection{Behaviour of $K^{*}(t)$ for $|\Delta T^\prime| \gg |\Delta T|$}

In addition to the limits $|\Delta T| \ll 1$ and $|\Delta T^{\prime}| \ll 1$, required by the linear analysis developed above, here the case in which $|\Delta T| \ll |\Delta T^{\prime}|$ will be addressed.
In this limit,  equation (\ref{2.13a}) gives
\begin{equation}\label{2.24}
    \phi_E(t_w) = x \simeq 1- \epsilon,
\end{equation}
where $\epsilon \equiv |\Delta T| / |\Delta T^{\prime}|$. This means that $\phi_{E}(t)$ is still very close to unity for $t=t_{w}$, i.e. $t_{w}$ is much smaller than the characteristic relaxation time of the energy at temperature $T$.
The Kovacs hump, given by  (\ref{2.15}), can be written as a function of $s=t-t_{w}$, $t_{w}$, and $\epsilon$, as
\begin{eqnarray}
    K^*(s)&= \left( 1 + \frac{1}{\epsilon} \right) \phi_{E} (s+t_{w} ) - \frac{\phi_{E}(s)}{\epsilon} \nonumber \\
    &= \phi_E(s)+
    \frac{t_w}{\epsilon}\phi_E^{\prime}(s)
    +\Or\left( t_{w}\right)+\Or\left(
    \frac{t_w^2}{\epsilon}\right)\, . \label{2.26}
\end{eqnarray}
As a  function of $s$, $K^{*}$ has a has a maximum at $s=s_k$ that, keeping only the first two terms on the right hand side of  (\ref{2.26}), is given by the solution of the equation
\begin{equation}\label{2.27}
    D_2(s_k)=\frac{\epsilon}{t_w}  \simeq \frac{1-\phi_E(t_w)}{t_w} \, .
\end{equation}
The above expression implies that $s_k$ is an increasing function of $t_w$ or, equivalently, it is a decreasing function of $|\Delta T^\prime|$, for $\Delta T$ fixed. This can be seen by taking the derivative with respect to $t_{w}$ in both sides of  (\ref{2.27}) to get
\begin{equation}\label{2.27b}
    D_2^{\prime}(s_k) \frac{ \partial s_k}{ \partial t_w}=\frac{\phi_E(t_w)-1-t_{w} \phi_E^{\prime}(t_w)}{t_w^{2}}
    \, .
\end{equation}
The right hand side of this equation  is negative, because the numerator is a strictly decreasing function of $t_{w}$  vanishing for $t_w\rightarrow 0$.  Moreover, $D_2$ was shown to be a monotonically decreasing function of time, following that $\partial s_{k}/\partial t_{w}>0$.

Let us analyze  equation (\ref{2.26}) deeper. Two rather different scenarios can arise, depending on the behaviour of $t_{w}$ with $\epsilon$, when the latter is very small. Consider first that to lowest order it is $t_{w} \sim \Or (\epsilon)$. For very short times, it is
\begin{equation}
\label{2.27d}
\phi_{E}(t) \simeq 1- D_{1}(0) t,
\end{equation}
where it has been used that $D_{1}(0) =- \phi^{\prime}(0)$. Particularization of the above relation for $t=t_{w}$ and use of  (\ref{2.24}) yields
\begin{equation}
\label{2.27dd}
t_{w}= \frac{\epsilon}{D_{1}(0)}.
\end{equation}
Substitution of this value into  (\ref{2.26}) leads to
\begin{equation}\label{2.26b}
    K^*(s)=\phi_E(s)+\frac{\phi_E^{\prime}(s)}{D_{1}(0)}\,+\Or(\epsilon)
    \, .
\end{equation}
Therefore, when $t_{w} \sim \Or(\epsilon)$, the maximum of the Kovacs function is located at a value $s_{k}^{(0)}$ verifying
\begin{equation}
\label{2.26bb}
D_{2} (s_{k}^{(0)}) = \frac{\epsilon}{t_{w}} \simeq D_{1}(0).
\end{equation}
Consequently, $s_{k}$ tends to a  finite, well defined value for $t_{w} \rightarrow 0$. This is consistent with the behaviour observed in some models, where $s_k$ seems to be almost independent of the magnitude of the second temperature jump \cite{TyV04,AAyN08}.

Consider next the case $t_{w} \ll \epsilon$. This may happen in systems where there is a time window for which $t \gg \left[ D_{1}(0)\right]^{-1}$, so that (\ref{2.27d}) does not apply, but $\phi_{E}(t)$ is still very close to unity and (\ref{2.24}) can be accomplished inside that window. A well known simple model exhibiting this behaviour is the one-dimensional Ising model with Glauber dynamics \cite{Gl63}  in the low temperature region \cite{ByP93,ByP96}. A typical time dependence in this time window is
\begin{equation}
\label{2.26cc}
\phi_{E}(t) \simeq 1 - A t^{\gamma},
\end{equation}
where $A>0$ is a constant and $\gamma$ a real parameter in the interval $0< \gamma < 1$. Then,  equation (\ref{2.24}) gives
\begin{equation}
\label{2.26dd}
t_{w} = \left( \frac{ \epsilon}{A} \right)^{1/\gamma}
\end{equation}
and  equation (\ref{2.26}) takes the form
\begin{equation}
K^{*}(s) = \phi_{E} (s)+ \Or \left( \epsilon^{-1+\frac{1}{\gamma}} \right).
\end{equation}
The above discussion supports the following scenario: as $|\Delta T^{\prime}|$ increases, $K^{*}(s)$ approaches the relaxation curve $\phi_{E}(s)$ because the maximum of the former $s_{k}$ moves to smaller times. It is interesting
to note that the stretched exponential
\begin{equation}
\label{2.26ee}
\phi_{KWW}(t) = e^{-\left(t/\tau \right)^{\gamma}},
\end{equation}
with $\tau$ being a characteristic relaxation time and $0 < \gamma < 1$, which is often used to fit experimental data of the relaxation of supercooled liquids and other complex systems \cite{Sch86,ANMMyM00}, has the short time behaviour given by  (\ref{2.26cc}).

\subsection{Behaviour of $K^{*}(t)$ for $|\Delta T^\prime| \ll |\Delta T|$}

For $|\Delta T^\prime| \ll |\Delta T|$,  (\ref{2.13}) leads to
\begin{equation}\label{2.38}
    \frac{|\Delta T^{\prime}|}{|\Delta T|}=\frac{x}{1-x}\ll 1,
\end{equation}
i.e. $x = \phi_{E}(t_{w}) \ll 1$ and, therefore, $t_{w} \gg 1$. As the equilibrium energies at the temperatures $T$ and $T+\Delta T^\prime=T_1$ are much closer than those for $T$ and $T- \Delta T=T_0$, the  curve describing the relaxation of the energy at temperature $T_{1}$ crosses $\langle E\rangle_e(T)$ at a later stage of the relaxation, as compared with
the case $|\Delta T| \ll |\Delta T^{\prime}|$ .

The function $D_1(t)$, introduced in  (\ref{2.15b1}), tends to a well defined limit $D_1^{(\infty)}$, as indicated in  (\ref{2.15c}). Therefore, $D_1^{\prime}(t)$ vanishes as $t \rightarrow\infty$ and this allows the approximation
\begin{equation}\label{2.40}
    \phi_E(t_w+s)=e^{\ln\phi_E(t_w)-D_1(t_w)s-
    \frac{1}{2}D_1^{\prime}(t_w)s^{2}+\cdots} \simeq
    x e^{-D_1(t_w)s} \, ,
\end{equation}
expected to be valid for $D_1^{\prime}(t_w)s^{2}\ll 1$. Thus, equation (\ref{2.15}) gives
\begin{equation}\label{2.39}
    K^*(s) \simeq \frac{1}{1-x} \left[ x\, e^{-D_1(t_w)s}-x\,\phi_E(s) \right] \simeq x \left[e^{-D_1(t_w)s}-\phi_E(s) \right].
\end{equation}
This function exhibits a maximum at  $s=s_k$ verifying
\begin{equation}\label{2.41}
    \phi_E^{\prime}(s_k)=-D_1(t_w)e^{-D_1(t_w)s_k} \, .
\end{equation}
Two  possibilities must be considered at this point, corresponding to a vanishing and non-vanishing long time limit,
$D_{1}^{(\infty)}$, of $D_{1}(t)$, respectively (see  (\ref{2.15c})). If $D_1^{(\infty)}=0$, it is $D_{1}(t_{w}) \ll 1$ and  equation (\ref{2.41}) can be approximated by
\begin{equation}\label{2.41b}
    \phi_E^{\prime}(s_k) \sim -D_1 (t_w) \equiv \frac{\phi_E^{\prime}(t_w)}{\phi_E(t_w)}\, ,
\end{equation}
as long as $D_{1}(t_{w}) s_{k} \ll 1$, something to be checked a posteriori. Since $\phi_{E}(t_{w}) \ll 1$, it is concluded from the above relation that $\phi^{\prime}_{E}(s_{k})  \gg  \phi^{\prime}_{E} (t_{w})$ and hence
$s_{k} \ll t_{w}$. Therefore, $s_{k}$ increases as $t_{w}$ increases and it diverges for $t_{w} \rightarrow \infty$,
but remaining always much smaller than $t_{w}$, i.e. $t_w \gg s_k \gg 1$. Moreover,  equation (\ref{2.41b}) also implies that
\begin{equation}
\label{2.41bb}
s_{k} D_{1}(t_{w}) \simeq -s_{k} \phi^{\prime}_{E}(s_{k}),
\end{equation}
and the right hand side of this relation is very small since $\phi_{E}(s)$ tends to zero when $s$ goes to infinity. This proves the consistency of the assumed dominant balance used to solve  (\ref{2.41}). The maximum value of the Kovacs function is obtained by substituting $s_{k}$ into  (\ref{2.39}),
\begin{equation}\label{2.43}
    K^{*}_{\text{max}} \equiv K^{*}(s_k)=x \left[ e^{-D_1(t_w)s_k}-\phi_E(s_k) \right] \sim x \, .
\end{equation}
Summarizing: for $|\Delta T^{\prime}| \ll |\Delta T|$ and $D_{1}^{(\infty)} =0$, the position of the maximum $s_{k}$ decreases as $|\Delta T^{\prime}|$ increases, while the height of the maximum $K^{*}_{\text{max}} \sim \phi_{E}(t_{w})$ has the opposite behaviour. This agrees with the experimental observations by Kovacs \cite{Ko63,Ko79}. It is worth stressing that the stretched exponential  (\ref{2.26ee}), verifies that  $D_1(t)\rightarrow 0$ for $t\rightarrow\infty$, and therefore it fits into the case just discussed.

Consider next that  $D_1^{(\infty)}>0$, so that  equation (\ref{2.39}) reduces to
\begin{equation}\label{2.56b}
    K^*(s) = x \left[ e^{-D_1^{(\infty)}s}-\phi_E(s)\right]=x \varphi(s) \, ,
\end{equation}
where $ \varphi(s) \equiv \exp [-D_{1}^{(\infty)}s] - \phi_{E}(s)$. Therefore, $K^*(s)$ factorizes into  a function of $t_w$ times a function of $s$. This means that the maximum of $K^{*}(s)$ will occur at a time $s_k^{(\infty)}$ such that $\varphi^\prime(s)$ vanishes, and it is independent of $t_w$,
\begin{equation}\label{2.56c}
    \phi_E^\prime(s_k^{(\infty)})=-D_1^{(\infty)}e^{-D_1^{(\infty)}s_k^{(\infty)}} \, ,
\end{equation}
i.e. $s_k^{(\infty)}$ is the formal limit of $s_k$ given by  (\ref{2.41}) for $t_w\rightarrow\infty$. The height of the maximum in this case is
\begin{equation}\label{2.56d}
    K_{\text{max}}^{*}=x \left[ e^{-D_1^{(\infty)}s_k^{(\infty)}}-\phi_E(s_k^{(\infty)})\right]\, .
\end{equation}
It is again a decreasing function of the waiting time $t_w$, or an increasing function of the second temperature jump $|\Delta T^\prime|$,  in agreement with the experimental observations. On the other hand, the position of the maximum, $s_k^{(\infty)}$, now does not diverge but tends to a finite limit. Also this is consistent with the behaviour found in some models, in which the position of the maximum of the hump seems to be independent of the second temperature jump \cite{TyV04,AAyN08}.

\subsection{Asymptotic behaviour of $K^{*}(t)$ for long times}

Let us start our analysis from  (\ref{2.15}) written in the form
\begin{equation}\label{2.57}
    K^{*}(s)=\phi_E(s)+\frac{1}{1-x}\left[ \phi_E(t_w+s)-\phi_E(s) \right] \, .
\end{equation}
For long enough times, $\phi_E(t_w+s)$ can be approximated by
\begin{equation}\label{2.58}
    \phi_E(t_w+s)=\phi_E(s) e^{-D_1(s)t_w} \, ,
\end{equation}
valid for $D_1^\prime(s)t_w^2\ll 1 $. Remember that $D_1^\prime(s)\rightarrow 0$ for $s\rightarrow\infty$. Use of  (\ref{2.58}) into   (\ref{2.57}) gives
\begin{equation}\label{2.59}
    K(s)\sim \phi_E(s)+\frac{\phi_E(s)}{1-x}\, \left[e^{-D_1(s)t_w} -1 \right]  \, ,
\end{equation}
for $s \gg 1$. As in the previous section, two cases must be analyzed separately. If  $D_{1}(s) \rightarrow 0$ in the long time limit, the above expression simplifies to
\begin{equation}\label{2.60}
    K(s) \sim \phi_E(s) \, ,
\end{equation}
 i.e., the Kovacs function $K^*(s)$ approaches the relaxation function $\phi_E(s)$ for long times, such that $D_1(s) t_w\ll 1$. This is analogous to the experimental observation \cite{Ko63,Ko79}. On the other hand, if the logarithmic derivative does not vanish in the long time limit but $D_1(t)\rightarrow D_1^{(\infty)}> 0$,  (\ref{2.59}) takes the form
\begin{equation}\label{2.62}
    K^{*}(s) \sim \phi_E(s) \frac{e^{-D_1^{(\infty)}t_w}-\phi_E(t_w)}{1-\phi_E(t_w)} \,
\end{equation}
or
\begin{equation}\label{2.63}
    \ln K(s) \sim \ln\phi_E(s) +\ln \frac{e^{-D_1^{(\infty)}t_w}-\phi_E(t_w)}{1-\phi_E(t_w)} \, ,
\end{equation}
for $s \gg 1$. Consequently, for long enough times plots of $\ln K(s)$ versus $s$ corresponding to different values of $t_w$ can be collapsed  on $\ln\phi_E(s)$, by subtracting an adequate constant quantity for each value of $t_w$.

\section{A two-level system with dynamical disorder}
\label{s3}

\begin{figure}
\begin{center}
\epsfig{file=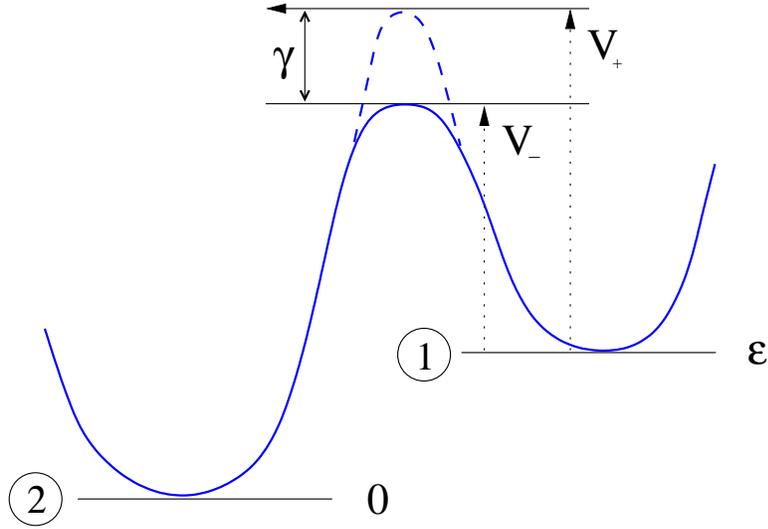, width=4in}
\caption[]{Schematic representation of the TLS with dynamical disorder described in the text.\label{fig2}}
\end{center}
\end{figure}

Here the general scenario developed in the previous sections will be particularized for a simple model, perhaps the simplest one exhibiting the Kovacs effect. It is a two-level system (TLS) with dynamical disorder. There are two possible states of the system that will be denoted by $1$ and $2$, respectively. The difference of energy between the states is $\varepsilon$, and the energy barrier between them, measured from the excited state $1$, is $V$. This barrier $V$ is not fixed but fluctuates in time between two values, $V_{+}$ and $V_{-}$, being $V_{+}>V_{-}$. These fluctuations are described by a dichotomic Markov process with constant transition rate $\gamma$. A sketch of the model is presented in figure \ref{fig2}.

The transition rates between states $1$ and $2$ are given by
\numparts
\begin{eqnarray}
\label{3.1a}
    W_{21}^{\pm}&=&W(1\rightarrow 2)^\pm=\alpha e^{-\beta V_\pm} \equiv\nu_\pm \, ,
\\
\label{3.1b}
    W_{12}^\pm &=&W(2\rightarrow 1)^\pm=\alpha e^{-\beta (V_\pm+\varepsilon)}=\nu_\pm e^{-\beta\varepsilon} \, ,
\end{eqnarray}
\endnumparts
where $\alpha$ is a characteristic attempt rate to cross the barrier and the $\pm$ signs correspond to the two possible values of the energy barrier. Let $\mathcal{P}_{\nu}(i,t)$ denote the probability of finding the system in state $i=1,2$ at time $t$, being in addition the height of the barrier $V_{\nu} ;\, \nu=\pm$. The master equation for the model reads
\begin{equation}\label{3.5}
    \frac{\partial}{\partial t}\mathcal{P}(t)=\mathbb{W}\mathcal{P}(t)
\end{equation}
with
\begin{equation}\label{3.6a}
    \mathcal{P} \equiv \left( \begin{array}{c}
    \mathcal{P}_+(1,t) \\ \mathcal{P}_+(2,t) \\ \mathcal{P}_-(1,t) \\ \mathcal{P}_-(2,t)
    \end{array} \right) \, ,
\end{equation}
and
\begin{equation}\label{3.6b}
    \mathbb{W}=\left(  \begin{array}{cccc}
    -(\nu_+ +\gamma) & \nu_+ e^{-\beta\varepsilon} & \gamma & 0 \\
    \nu_+ & -(\nu_+ e^{-\beta\varepsilon}+\gamma) & 0 & \gamma \\
    \gamma & 0 & -(\nu_- +\gamma) & \nu_-e^{-\beta\varepsilon} \\
    0 & \gamma & \nu_- & -(\nu_-e^{-\beta\varepsilon} +\gamma)\,
    \end{array}  \right).
  \end{equation}

The equilibrium distribution $\mathcal{P}_e$ is the stationary solution of  (\ref{3.5}),
\begin{equation}\label{3.7}
    \mathcal{P}_e=\frac{1}{2(1+\rho)}\left(  \begin{array}{c}
    \rho \\ 1 \\ \rho \\ 1
    \end{array}  \right),
\end{equation}
where we have introduced the parameter
\begin{equation}\label{3.8}
    \rho \equiv e^{-\beta\varepsilon} \, .
\end{equation}
Note that the matrix $\mathbb{W}$ in  (\ref{3.6b}) satisfies the detailed balance condition.
The eigenvalues $-\lambda^{(\alpha)}$ and eigenvectors $\psi^{(\alpha)}$ of $\mathbb{W}$ are the solutions of the equation
\begin{equation}
\label{3.11}
    \mathbb{W}\psi^{(\alpha)}=-\lambda^{(\alpha)}\psi^{(\alpha)} \, .
\end{equation}
In addition to the null eigenvalue corresponding to the equilibrium distribution, there are other three eigenvalues given by
\begin{equation}
\label{3.12a}
    \lambda^{(1)}=\gamma+\nu_M(1+\rho)-\kappa \, ,
\end{equation}
\begin{equation}
\label{3.12b}
    \lambda^{(2)}=\gamma+\nu_M(1+\rho)+\kappa \, ,
\end{equation}
\begin{equation}
\label{3.12c}
\lambda^{(3)}=2\gamma \, .
\end{equation}
The associated eigenvectors are
\begin{equation}
\label{3.12aa}
\psi^{(1)}=\sqrt{\frac{\rho}{2(1+\rho)^2}}
\frac{\gamma}{\sqrt{[\kappa+\delta(1+\rho)]^2+\gamma^2}}
\left(  \begin{array}{c}
-\frac{\kappa+\delta(1+\rho)}{\gamma} \\ \frac{\kappa+\delta(1+\rho)}{\gamma} \\ -1 \\ 1
\end{array}  \right),
\end{equation}
\begin{equation}
\label{3.12bb}
 \psi^{(2)}=\sqrt{\frac{\rho}{2(1+\rho)^2}}
\frac{\gamma}{\sqrt{[\kappa-\delta(1+\rho)]^2+\gamma^2}}
\left(  \begin{array}{c}
\frac{\kappa-\delta(1+\rho)}{\gamma} \\ \frac{\delta(1+\rho)-\kappa}{\gamma} \\ -1 \\ 1
\end{array}  \right),
\end{equation}
\begin{equation}
\label{3.12cc}
\psi^{(3)}=\frac{1}{2(1+\rho)}
\left(  \begin{array}{c}
-\rho \\ -1 \\ \rho \\ 1
\end{array}  \right)\, .
\end{equation}
In the above expressions, the positive parameters
\begin{equation}\label{3.9a}
    \nu_M \equiv \frac{\nu_- +\nu_+}{2} \, ,
\end{equation}
\begin{equation}\label{3.9b}
    \delta \equiv \frac{\nu_- -\nu_+}{2} \, ,
\end{equation}
\begin{equation}\label{3.10}
    \kappa \equiv \sqrt{\gamma^2+\delta^2 (1+\rho)^2} \, ,
\end{equation}
have been introduced.

For any two matrices $\mathcal{P}_{1}$ and $\mathcal{P}_{2}$ defining states of the system,  a scalar product is defined as \cite{vk92}
\begin{equation}
\label{3.10a}
\left( \mathcal{P}_{1} , \mathcal{P}_{2} \right) \equiv \sum_{i=1}^{4} \frac{\mathcal{P}_{1,i} \mathcal{P}_{2,i}}{ \mathcal{P}_{e,i}}.
\end{equation}
Then, the eigenvectors $\psi^{(\alpha)}$ verify the orthogonality condition
\begin{equation}\label{3.13}
    \left(\psi^{(\alpha)},\psi^{(\alpha^\prime)}\right)=\delta_{\alpha \alpha^\prime} \, .
\end{equation}
and they constitute an orthonormal basis for the solutions of the master equation,
\begin{equation}\label{3.14}
    \mathcal{P}(t)=\mathcal{P}_e+\sum_{\alpha=1}^{3} a^{(\alpha)} \psi^{(\alpha)} e^{-\lambda^{(\alpha)}t} \, .
\end{equation}
The coefficients $a^{(\alpha)}$ are given by
\begin{equation}\label{3.15}
    a^{(\alpha)}=\left( \psi^{(\alpha)},\mathcal{P}(0) \right)\, .
\end{equation}
In this way, the dynamics of the system is completely solved, since the evolution of the probability distribution $\mathcal{P}(t)$ has been obtained for an arbitrary initial condition $\mathcal{P}(0)$.

The marginal probability $\mathcal{P}_{\nu}(t)$ of finding the TLS with a barrier $V_{\nu}, \nu=\pm$, at time $t$,
regardless of the state of the system, is given by
\begin{equation}\label{3.16}
    \mathcal{P}_\nu(t)=\mathcal{P}_\nu(1,t)+\mathcal{P}_\nu(2,t) \, .
\end{equation}
Similarly,
\begin{equation}\label{3.17}
    \mathcal{P}(r,t)=\mathcal{P}_+(r,t)+\mathcal{P}_-(r,t) \,
\end{equation}
is the probability of finding the system in state $r=1,2$, without taking into account the value of the energy barrier. It is interesting to realize that
\begin{eqnarray}
    a^{(3)}=\left( \psi^{(3)},\mathcal{P}(0) \right) &=-\mathcal{P}_+(1,0)-\mathcal{P}_+(2,0)+\mathcal{P}_-(1,0)+\mathcal{P}_-(2,0)
    \nonumber \\
    &=\mathcal{P}_-(0)-\mathcal{P}_+(0) \, ,
\end{eqnarray}
i.e. $a^{(3)}=0$ if the initial probability is symmetrically distributed between the two possible values of the barrier. Thus the third eigenvalue $\lambda^{(3)}$ is associated with the relaxation towards configurations in which both values of $V$ have the same probability.

An expression for the equilibrium correlation function of the energy can be written down by particularizing  (\ref{2.7}),
\begin{equation}\label{3.19}
    \langle E(t)E(t^\prime) \rangle_e=\langle E\rangle_e^2+\sum_{\alpha=1}^{2} \left( \Delta E^{(\alpha)}\right)^2 e^{-\lambda^{(\alpha)}(t-t^\prime)} \, ,
\end{equation}
with $\Delta E ^{(\alpha)}$ given by  (\ref{2.7a}). The term corresponding to $\alpha=3$ has been omitted since it is $\Delta E^{(3)}=0$. Taking the origin of energies at the ground state $2$, it is
\begin{equation}\label{3.21}
    \Delta E^{(\alpha)}=\sum_i \varepsilon_i \psi_i^{(\alpha)}= \varepsilon \left[ \psi_1^{(\alpha)}+\psi_3^{(\alpha)} \right]
\end{equation}
or, explicitly,
\numparts
\begin{eqnarray}
\label{3.22a}
\Delta E^{(1)} & = &- \varepsilon \sqrt{\frac{\rho}{2(1+\rho)^2}}\frac{\kappa+\delta(1+\rho)+\gamma}{\sqrt{[\kappa+\delta(1+\rho)]^2+\gamma^2}}
\, , \\
\label{3.22b}
\Delta E^{(2)} & = & - \varepsilon \sqrt{\frac{\rho}{2(1+\rho)^2}}\frac{-\kappa+\delta(1+\rho)+\gamma}{\sqrt{[\kappa-\delta(1+\rho)]^2+\gamma^2}}
\, ,
\end{eqnarray}
\endnumparts
For the equilibrium energy dispersion it is found
\begin{equation}\label{3.23}
    \sigma_E^2= \left( \Delta E^{(1)}\right)^2+\left(\Delta E^{(2)}\right)^2= \frac{\varepsilon^2 \rho}{(1+\rho)^2}\, .
\end{equation}
The relaxation function of the energy follows by applying  (\ref{2.11}),
\begin{equation}\label{3.23b}
    \phi_E(t)=b^{(1)} e^{-\lambda^{(1)}t}+b^{(2)} e^{-\lambda^{(2)}t} \, ,
\end{equation}
with
\numparts
\begin{eqnarray}
\label{3.24a}
b^{(1)} & =\frac{1}{2} \frac{[\kappa+\delta(1+\rho)+\gamma]^2}{[\kappa+\delta(1+\rho)]^2+\gamma^2} \, ,
\\
\label{3.24b}
b^{(2)} & =\frac{1}{2} \frac{[-\kappa+\delta(1+\rho)+\gamma]^2}{[\kappa-\delta(1+\rho)]^2+\gamma^2} \, .
\end{eqnarray}
\endnumparts
The long time behaviour of $\phi_E(t)$ is given by
\begin{equation}\label{3.24c}
    \phi_E(t)\sim b^{(1)} e^{-\lambda^{(1)}t} \, , \quad t\rightarrow\infty \, ,
\end{equation}
since $\lambda^{(1)}<\lambda^{(2)}$. Therefore, the long time limit of the function $D_1(t)$ defined in  (\ref{2.15b1}) is
\begin{equation}\label{3.24d}
    D_1^{(\infty)}=\lambda^{(1)}>0 \, .
\end{equation}
This is a relevant feature of the present model since, as discussed in section \ref{s2}, the long time limit of $D_1(t)$ controls some of the key properties of the function $K^{*}(t_{w},s)$ characterizing  the Kovacs effect.

In the rapidly fluctuating barrier limit defined by $\gamma\gg\delta(1+\rho)$, the system is equivalent to a TLS with an effective barrier,
\numparts\begin{eqnarray}
\label{3.29a}
    \lambda^{(1)} \sim \nu_M(1+\rho) \, , & \quad \lambda^{(2)} \sim 2\gamma+\nu_M(1+\rho)\, ,\\
    \label{3.29b}
    b^{(1)} \sim 1\, , & \quad b^{(2)} \ll 1 .
\end{eqnarray}\endnumparts
Therefore, the relaxation approaches an exponential decay as $\gamma$ increases.

In the slowly fluctuating barrier limit, $\gamma\ll\delta(1+\rho )$, the static disorder case is recovered. The results tend to those of an ensemble of two kinds of  TLS  having the activation energy $\varepsilon$ but different barrier heights, $V_+$ and $V_-$, respectively,
\numparts\begin{eqnarray}
    \label{3.27a}
    \lambda^{(1)}&  \sim  \nu_+(1+\rho) \, , \\
\label{3.27b}    \lambda^{(2)}& \sim \nu_-(1+\rho) \, .
\end{eqnarray}\endnumparts
Besides,
\begin{equation}\label{3.28}
    b^{(1)} \sim \frac{1}{2}\, , \quad b^{(2)} \sim\frac{1}{2}\, .
\end{equation}

\section{The Kovacs  experiment in the disordered TLS}\label{s3a}

Next, the results of submitting a disordered TLS to the Kovacs experiment will be discussed. In the linear response approximation, valid for small temperature jumps, the function $K^{*}(t)$ given in  (\ref{2.15}) characterizes the evolution of the  energy after the quench from the intermediate temperature $T_{1}$ to the final temperature $T$, at time $t_w$. The latter is determined by  (\ref{2.13a}), that using  (\ref{3.23b}) becomes
\begin{equation}\label{3.31}
    b^{(1)}e^{-\lambda^{(1)}t_w}+b^{(2)}e^{-\lambda^{(2)}t_w}=
    \frac{|\Delta T^\prime|}{|\Delta T|+|\Delta  T^\prime|} \equiv x \, ,
\end{equation}
which gives $t_w$ as a function of the temperature jumps ratio $|\Delta T^\prime|/|\Delta T|$. Substituting  (\ref{3.23b}) into  (\ref{2.15}) leads after some algebra to
\begin{equation}\label{3.35b}
    K^{*}(t_{w},s)=K_0(t_w)K_1(s) \, ,
\end{equation}
where $s= t -t_{w}$,
\begin{equation}\label{3.36}
    K_0(t_w)=b^{(1)}b^{(2)} \frac{e^{-\lambda^{(1)}t_w}-e^{-\lambda^{(2)}t_w}}{1-\phi_E(t_w)}\, ,
\end{equation}
\begin{equation}\label{3.37}
    K_1(s)=e^{-\lambda^{(1)}s}-e^{-\lambda^{(2)}s} \, .
\end{equation}
A factorization similar to  (\ref{3.35b}) was found in the previous section for a general Markov system such that $D_1^{(\infty)}\neq 0$,  equation (\ref{2.56b}), but there it was restricted to the limit $| \Delta T^\prime|\ll\ |\Delta T|$ or, equivalently, large values of $t_w$. Here, it has been obtained as a general result for the disordered TLS in the linear response approximation.

The position of the maximum of $K^{*}$ as a function of $s$ does not depend on $t_w$, and it is given by
\begin{equation}\label{3.38}
    s_k=\frac{1}{\lambda^{(2)}- \lambda^{(1)}} \ln\frac{\lambda^{(2)}}{\lambda^{(1)}} \, .
\end{equation}
The maximum value, for given $t_{w}$, is
\begin{equation}\label{3.44}
    K^{*}_{\text{max}}=K_0(t_w)K_1(s_k)=b^{(1)}b^{(2)}y^{\frac{y}{1-y}}(1-y)
    \frac{e^{-\lambda^{(1)}t_w}-e^{-\lambda^{(2)}t_w}}{1-\phi_E(t_w)} \, ,
\end{equation}
where the parameter
\begin{equation}\label{3.39}
    y \equiv\frac{\lambda^{(1)}}{\lambda^{(2)}}<1 \,
\end{equation}
has been introduced. In the limit $t_w\rightarrow 0$ ($|\Delta T |\ll |\Delta T^{\prime}|$) it is
\begin{equation}\label{3.42}
    K_{0}(t_w)  \sim b^{(1)}b^{(2)} \frac{\lambda^{(2)}-\lambda^{(1)}}{b^{(1)}\lambda^{(1)}+b^{(2)}\lambda^{(2)}} \, ,
\end{equation}
 showing that the higher the Kovacs effect the further from a single exponential the shape of the relaxation. On the other hand, for $t_w\rightarrow\infty$ ($|\Delta T| \gg |\Delta T^{\prime}|$),
 \begin{equation}\label{3.43}
    K_0(t_w)\ \ll 1 \, .
\end{equation}
Note that given the independence of $s_{k}$ from $t_{w}$,  equation (\ref{3.42}) requires that $s_{k} \gg t_{w}$, while equation
 (\ref{3.43}) holds for $s_{k} \ll t_{w}$. This is due to the factorization of $K^{*}(s)$ and it seems to be also the case in other simple models \cite{TyV04,AAyN08}. Between the two above limits, $K_0(t_{w})$ decreases monotonically, since $K_0^\prime(t_w)<0$ for all $t_w$, as it can be easily checked by direct computation. Therefore,
\begin{equation}\label{3.46}
    \frac{\rmd K_{\text{max}}^*}{\rmd t_w}=K_0^\prime(t_w)K_1(s_k)<0 \, .
\end{equation}

\begin{figure}
\begin{center}
\epsfig{file=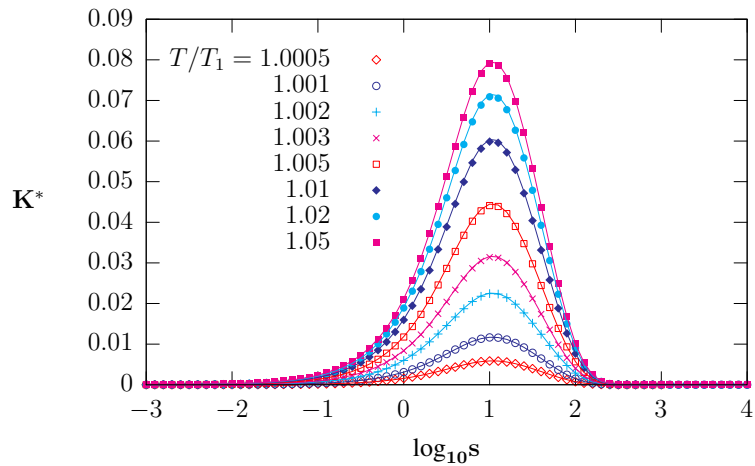, width=4in}
\caption[]{Plot of the dimensionless Kovacs function $K^{*}(t_w,s)$ as a function of the time after the second temperature jump $s$, for different values of the second temperature jump, and fixed values of the initial and final temperatures. The values of the parameters of the model, as well as the values of the fixed initial and final temperatures and the time unit used are given in the text. The symbols correspond to the numerical integration of the master equation while the solid lines are the theoretical predictions, given by (\protect{\ref{3.35b}}).  \label{fig3}}
\end{center}
\end{figure}

In figure \ref{fig3}, the Kovacs hump function $K^{*}$ is plotted as a function of time $s$ for different values of the intermediate temperature $T_1$, as indicated in the insert. For all the curves, $T=0.43$, $V_-=3$, $V_+=7$, and $\gamma=0.1$. The units of temperature, energy and time have been fixed by taking $k_B=1$, $\varepsilon=1$ and $\alpha=1$, respectively. The initial temperature is such that $T=0.999 T_0$. A very good agreement is found between the numerical solution of the master equation and the theoretical predictions, for all the cases considered. The curves have a maximum which is located at a fixed position, accurately  predicted by  (\ref{3.38}) which gives $s_k\approx 10.4$, i.e. $\log_{10} s_k\approx 1.04$. The height of the maximum is a decreasing function of $T_1$, i.e. an increasing function of the second temperature jump $|\Delta T^\prime|$, as implied by  (\ref{3.46}).

\begin{figure}
\begin{center}
\epsfig{file=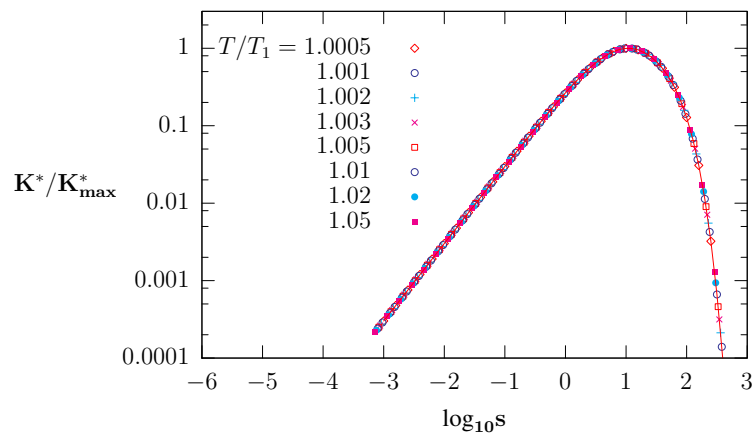, width=4in}
\caption[]{Plot of $K^{*}(t_{w},s)/K^*_{\text{max}}(t_w)$ as a function of time $s$, for the same values of the parameters as in figure \protect{\ref{fig3}}. Time is measured in units of the TLS attempt rate $\alpha$. Note the logarithmic scale in the vertical axis. The symbols correspond to the numerical integration of the master equation and the line to the analytical expression (\protect{\ref{3.44b}}). \label{fig4}}
\end{center}
\end{figure}

As a consequence of the factorization property given in  (\ref{3.35b}),
\begin{equation}\label{3.44b}
    \frac{K^{*}(t_{w},s)}{K^{*}_{\text{max}}(t_{w})}=\frac{K_1(s)}{K_1(s_k)} \, ,
\end{equation}
that is independent from $t_w$. This is verified in figure \ref{fig4}, where the same data as in figure \ref{fig3} are shown, but now scaling the values of $K^{*}(t_{w}, s)$ with $K^*_{\text{max}}(t_{w})$. It is seen that all the curves collapse as required by  (\ref{3.44b}).

\begin{figure}
\begin{center}
\epsfig{file=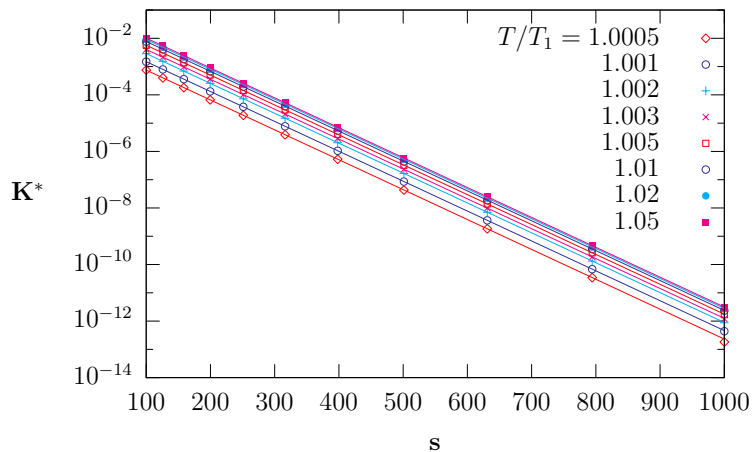, width=4in}
\caption[]{Plot of $K^{*}(t_{w},s)$ for the same systems as in figure \protect{\ref{fig3}}, in the long time region $100 \leq s\leq 1000$.  Time $s$ is measured in units of the TLS attempt rate $\alpha$. The symbols correspond to the numerical integration of the master equation, while the lines are the theoretical result  \protect(\ref{3.35b}).\label{fig5} }
\end{center}
\end{figure}

Now, we are going to study the long time behaviour of $K^*(s)$. From  (\ref{3.35b})-(\ref{3.37}), it is found that
for $t_{w} \rightarrow \infty$
\begin{equation}\label{3.54}
    \ln K^{*}(t_{w},s) \sim \ln K_0(t_w) -\lambda^{(1)}s \, , \quad s\rightarrow\infty \, ,
\end{equation}
or, taking into account the form of the long time limit of $\phi_E(s)$  (\ref{3.24c}),
\begin{equation}\label{3.54b}
    \ln K^{*}(t_{w},s) \sim \ln\phi_E(s)+\ln \left[  \frac{K_0(t_w)}{b^{(1)}} \right] \, .
\end{equation}
The difference between $\ln K^{*}(t_{w},s)$ and $\ln\phi_E(s)$ becomes independent of $s$ for $s\rightarrow\infty$, consistently with  (\ref{2.63}). In figure \ref{fig5}, the Kovacs function is plotted, for the same values of the parameters as in figure \ref{fig3}, in the long time window $10^{2} \leq s\leq 10^{3}$. All the curves are straight lines of slope $-\lambda^{(1)}$, as predicted by  (\ref{3.54}). It is also seen that for small values of the waiting time, corresponding in the figure to $T/T_{1}>1.02$, the long time behaviour of $K^{*}(t_{w},s)$ becomes independent of $t_w$, as it follows from (\ref{3.42}).

For high $\gamma$, namely $\gamma\gg 2 \delta(1+\rho)$, the parameter $\kappa$ defined in  (\ref{3.10}) behaves as
\begin{equation}\label{3.60}
    \kappa \sim \gamma+\frac{\delta^2 (1+\rho)^2}{2\gamma}\, ,
\end{equation}
while the corresponding limits for $\lambda^{(1)}$, $\lambda^{(2)}$, $b^{(1)}$ and $b^{(2)}$ are given by (\ref{3.29a}) and (\ref{3.29b}). Moreover, these limits imply that the  relaxation function $\phi_E(t)$ tends to an exponential, and, consequently, the  Kovacs effect decreases. In particular, taking into account that $y=\lambda^{(1)}/\lambda^{(2)}\rightarrow 0$, it follows from  (\ref{3.44}) that   $K^{*}_{\text{max}}$ also goes to zero.

\begin{figure}
\begin{center}
\epsfig{file=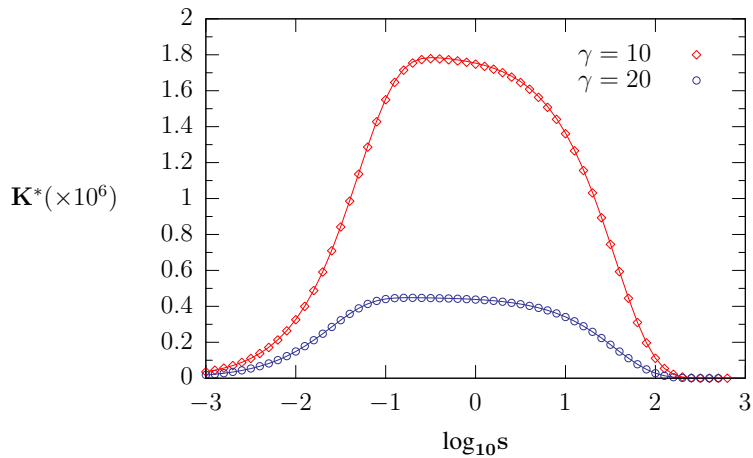, width=4in}
\caption[]{Plot of $K^{*}$, for the same values of the parameters of the model of figure \protect{\ref{fig3}}, for large $\gamma$, i.e. rapid barrier fluctuations. Units are the same as in the previous figures. The symbols correspond to the numerical integration of the master equation, while the lines are the theoretical expression \protect(\ref{3.35b}). Note that $K^{*}$ is very small.\label{fig7} }
\end{center}
\end{figure}

This is illustrated in figure \ref{fig7}, where the Kovacs function is plotted for two different high values of $\gamma$. The rest of the parameters are the same as in figure \ref{fig1}. The temperatures values are $T=0.999 T_0=1.001 T_1$. The height of the maximum decreases with $\gamma$, while its width increases. The increase of the width can be understood by calculating the second derivative of $K^*$ at the maximum. A straightforward calculation yields
\begin{equation}\label{3.61b}
    K^{*\prime\prime}(s_k)=K_0(t_w)K_1^{\prime\prime}(s_k)=
    K_0(t_w) \lambda^{(2)^2} y^{\frac{y}{1-y}} (y^2-y) \, ,
\end{equation}
As $y\rightarrow 0$ in the  limit of large $\gamma$, it follows that
\begin{equation}\label{3.62}
    K^{*\prime\prime}(s_k)=-b^{(1)}b^{(2)}\lambda^{(2)^2} \frac{e^{-\lambda^{(1)}t_w}-e^{-\lambda^{(2)}t_w}}{1-\phi_E(t_w)} y^{\frac{1}{1-y}} \rightarrow 0 \, ,
\end{equation}
since $b^{(2)}\lambda^{(2)^2}$ is seen to go to a constant by making use of (\ref{3.12b}), (\ref{3.24b}) and (\ref{3.60}). Therefore, the maximum becomes flatter as $\gamma$ increases, in agreement with the behaviour shown in figure \ref{fig7}.

\begin{figure}
\begin{center}
\epsfig{file=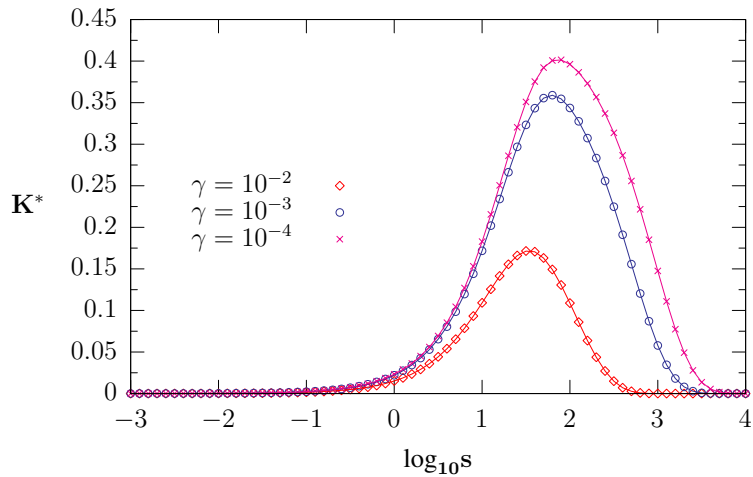, width=4in}
\caption[]{The Kovacs function $K^*$, for the same values of the parameters as in  figure \protect{\ref{fig7}}, except that now $\gamma$ is chosen in the regime of slow fluctuations of the barrier, as discussed in the main text. The symbols correspond to the numerical integration of the master equation, while the lines are the theoretical expression  \protect(\ref{3.35b}).\label{fig6} }
\end{center}
\end{figure}

Consider now the regime of slow barrier fluctuations, $ \gamma\ll 2 \delta(1+\rho)$, for which $\kappa$ becomes
\begin{equation}\label{3.58}
    \kappa \simeq \delta(1+\rho)\, ,
\end{equation}
and the corresponding behaviours of $\lambda^{(1)}$, $\lambda^{(2)}$, $b^{(1)}$ and $b^{(2)}$ are given by (\ref{3.27a}), (\ref{3.27b}), and (\ref{3.28}). As discussed around those equations, the static disorder limit is recovered and the nonexponential character of the relaxation is maximal. Therefore, the Kovacs effect is expected to increase as $\gamma$ decreases, reaching a maximum in the static disorder limit. This is clearly observed in figure \ref{fig6}, where $K^{*}(t_{w},s)$ is plotted for the same values of the other parameters as in figure \ref{fig7}, except that  three small values of $\gamma$ are considered now. The curve corresponding to the lowest value of $\gamma$, namely $\gamma=10^{-4}$ is indistinguishable from the results for the static disorder. The above discussion supports the idea that the Kovacs effect can be understood as a measure of the non-exponential character of the relaxation, in the sense that the height of the peak increases and its width decreases as the relaxation function separates from the exponential decay.

\begin{figure}
\begin{center}
\epsfig{file=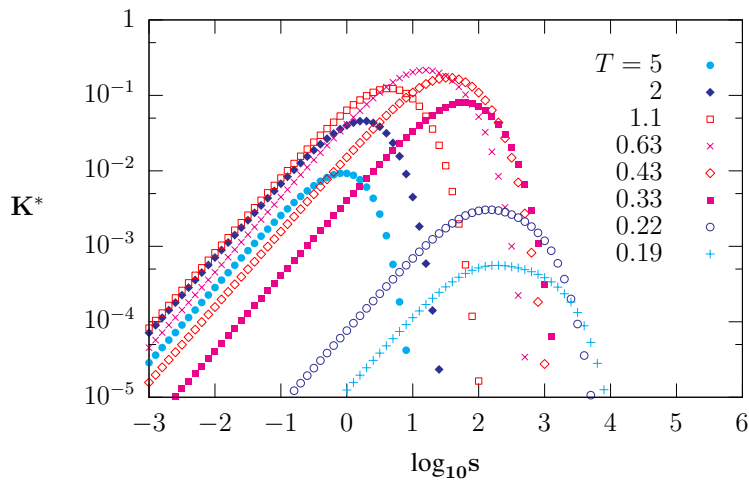, width=4in}
\caption[]{The relaxation function $K^{*}$ for the same values of the parameters of the model of the previous figures, but varying the value of the final temperature $T$. The values of the ratio between the  initial and intermediate temperatures have been kept constant, and they are given in the main text. The Kovacs effect shows a non-monotonic behaviour, being maximal for $T \approx 0.63$. The symbols correspond to the numerical integration of the master equation. For the sake of clarity, the theoretical lines have not been plotted.\label{fig8}
}
\end{center}
\end{figure}

Finally, the variation of the Kovacs effect with the temperature is shown in figure \ref{fig8}, where the dynamical disorder rate is $\gamma=0.01$ in all cases. The temperature jumps are the same as in Figs.\  \ref{fig7} and \ref{fig6}, $ T=0.999 T_{0}=1.001 T_{1}$, but several values of the final temperature $T$ have been considered. The height of the maximum $K^{*}_{\text{max}}$ as a function of $T$ exhibits a non-monotonic behaviour, with a maximum for $T \approx 0.63$. This ``resonant'' Kovacs behaviour can be easily understood.
Both in the high and the low temperature regimes, the relaxation is almost exponential. The reason is that the rapidly fluctuating barrier  condition $\gamma\gg\delta(1+\rho)$ is verified for $T\rightarrow\infty$ and also for $T\rightarrow 0$, because $\delta\rightarrow 0$ in both limits. For high temperatures, it is  $V_-,V_+ \ll k_B T$, and  from (\ref{3.1a}) and (\ref{3.1b}) it follows that $\nu_-\simeq \nu_+ \rightarrow \alpha$, i.e. the difference between the barriers become negligible, and the system is equivalent to a TLS with an average barrier. The situation in the low temperature regime is more subtle. In the limit $k_B T\ll V_-,V_+$, it is $\nu_-\ll\nu_+$ and, for instance, a system of two independent TLS with those parameters should exhibit a strong nonexponential behaviour. But in the system with dynamical disorder, when both rates $\nu_-$ and $\nu_+$ tend to zero while $\gamma$ reamins constant, the fluctuations of the barrier are much more rapid than the transitions over it and, consequently, the system is again equivalent to a TLS with an average barrier. This means that $K^*_{\text{max}}$ will show a maximum as a function of the temperature when $\gamma$ is of the order of the difference between the characteristic rates of the TLS with barriers $V_-$ and $V_+$. Then the maximum must occurs roughly at a  the temperature $\widetilde{T}$ such that
\begin{equation}\label{3.63}
    \gamma \approx 2 \delta (1+\rho) \, .
\end{equation}
For the values of the parameters of figure \ref{fig8}, $\widetilde{T}\approx 0.627$, in very good agreement with the reported numerical data.

How good remains the linear response approximation as the $T$ jumps are increased? For the range of parameters considered in this section, the agreement between the theoretical and the numerical integration curves remains good up to $|\Delta T|/T$ and $|\Delta T^\prime|/T$ of about $0.01$. Here, good means that the difference between both curves is not seen over the scale of the figures. If one of the jumps is further increased, multiplying it by a factor, the other has to be accordingly decreased, dividing it by the same factor, in order to keep the relative error roughly constant.

\section{Conclusions}
\label{s4}

In this paper, the Kovacs or crossover effect has been analyzed for systems whose dynamics is described by a master equation. The temperature jumps have been assumed to be small enough so as to apply the linear response approximation. This has allowed to write down a general expression for the relaxation function $K^{*}$ describing the Kovacs effect, in terms of the linear relaxation function $\phi_E$ and the ratio of the temperature jumps. The structure of this general expression  (\ref{2.15}), is similar to the one found by Kovacs in his phenomenological description \cite{Ko79}. The main difference lies in the arguments of the linear relaxation function $\phi_E$. While here they are real times, in Kovacs' expressions they are some effective times, in the spirit of the Narayanaswami-Moynihan-Tool phenomenological theory of supercooled liquids \cite{Na71,BEMyM76,To46}. Interestingly, these effective times can be seen to reduce to real times if the temperature jumps are considered to be small enough.

Some general properties of the function $K^{*}$ have been proved. They can be summarized as follows. The function $K^{*}$ is bounded, $0\leq K^{*}(t)\leq \phi_E(t)\leq \phi_E(s)$, with $s=t-t_w$. Besides, the behaviours of $K^*(s)$ in the limits $|\Delta T^\prime|\gg |\Delta T|$, $| \Delta T^\prime|\ll |\Delta T|$, and $t\rightarrow\infty$ are controlled by the behaviour of the logarithmic derivative of the direct relaxation function $\phi_E(t)$, i.e. of the function $D_1(t)$ introduced in  (\ref{2.15b}). If $D_1(t_w)$ has well defined limits for $t_w\rightarrow 0$ and $t_w\rightarrow\infty$, different from both zero and infinity, the position of the maximum $s_k$ has also well defined limits in both cases. On the other hand, if $D_1(0)\rightarrow\infty$, the position of the maximum $s_k$ goes to zero for $t_w\rightarrow 0$, i.e. when $|\Delta T^\prime|\gg|\Delta T|$. Besides, if $D_1(\infty)\rightarrow 0$, $s_k$ diverges for $t_w\rightarrow\infty$ . In general, the amplitude of the maximum increases with the second temperature jump $|\Delta T^\prime|$. For very long times, $K^{*}(s)$ tends asymptotically to $\phi_E(s)$ if the long time limit of $D_1$ vanishes. In this case, the direct relaxation decay, without second temperature jump, is recovered. If $D_1(\infty)\neq 0$, the recovery of the direct relaxation behaviour happens  in a weaker sense.  It can only be assured that $\ln K^{*}(s)-\ln\phi_E(s)$ becomes a function of the temperature jumps ratio $|\Delta T^\prime|/|\Delta T|$ for large $s$. Then, all the curves of $\ln K^{*}(s)$ corresponding to different second jumps, but to the same initial and final temperatures, can be collapsed on the curve $\ln\phi_E(s)$ by subtracting an adequate constant to each of them.

The experimental observations reported by  Kovacs \cite{Ko63,Ko79} are fully consistent  with the picture discussed above for the case in which $D_1(0)\rightarrow\infty$ and $D_1(\infty)\rightarrow 0$. This is reasonable, since the the KWW relaxation, often found in supercooled and other slowly relaxing systems, is also associated to this behaviour of the logarithmic derivative of the relaxation funtion. An important conclusion of the analysis carried out here is that the Kovacs effect is still present if the temperature jumps are small, i.e. the experimentally observed features persist in the linear response regime. The use of the detailed balance condition in the analysis developed here deserves some comments. This condition is actually necessary only to assure that the relaxation function of the energy is a monotonically decreasing function of time, namely a linear combination of exponentials with all the coefficients being positive. But the main results of section \ref{s2} remain valid if detailed balance does not hold, but the right eigenvectors of the transition rates matrix $\mathbb{W}$ are still a basis for the solutions of the master equation. Then, making use of both the left and right eigenvectors of $\mathbb{W}$, the relaxation function $\phi_E$ would be again a linear combination of exponentials. The problem is that, for this situation, the coefficients of the modes are not necessarily positive. But, in those cases in which they are,  $\phi_E$ will present a monotonic decay as a function of time, as given by  (\ref{2.11}), and all the subsequent results of Secs.\ \ref{s2} and \ref{s2a}  will remain valid.

As an application of the general theory, a simple model has been considered. It is a two-level system with dynamical disorder. For this model, it has been possible to obtain  an analytical closed equation for $K^{*}(t_{w},s)$, which is found to factorize, $K^{*}(t_{w},s)=K_0(t_w)K_1(s)$. As a consequence, the position of the maximum $s_k$ does not depend on $t_w$, i.e. on the second temperature jump. This is consistent with the behaviour of $D_1$, which has finite, nonzero, limits for both $t\rightarrow 0$ and $t\rightarrow\infty$. The independence of $s_k$ on the second temperature jump has also been observed in other models \cite{TyV04,AAyN08}. The height of the maximum, $K^*_{\text{max}}$, of $K^{*}$ is an increasing function of $|\Delta T^{\prime}|$ in its whole range of variation, since $K_0^\prime(t_w)<0$. Due to the factorization, all the curves corresponding to the same initial and final temperatures, but different second temperature jumps, collapse if they are rescaled with $K^*_{\text{max}}$. The dependence of the Kovacs effect on the dynamical disorder has also been investigated, for fixed temperatures. For a rapidly fluctuating barrier, the Kovacs effect disappears, because $K^*_{\text{max}}\rightarrow 0$. This is logical, since our system is equivalent to a TLS with an average barrier in this regime, and the relaxation is exponential. On the contrary, for a slowly fluctuating barrier, the Kovacs effect is maximal and the static disorder limit is recovered. The model becomes equivalent to two independent TLS with different barriers. Finally, we have also analyzed the dependence of the Kovacs effect on the final temperature, keeping constant the relative magnitude of the jumps and the dynamical disorder. Interestingly, a non-monotonic dependence of $K^*_{\text{max}}$ on the temperature shows up. This is a resonance like phenomenon, since the temperature for which $K^*_{\text{max}}$ is maximal is the one at which two characteristic rates of the system are of the same order, as given by  (\ref{3.63}).

\ack

This research was partially supported by the Ministerio de
Educaci\'{o}n y Ciencia (Spain) through Grant No. FIS2008-01339 (partially financed
by FEDER funds).

\appendix

\section{Linear response function at equilibrium}
\label{apa}

Here the linear response function $R_{A}(t,t^{\prime})$, defined in  (\ref{1.25}), will be particularized for an initial equilibrium state, $\mathcal{P}_{e,i}(T)$. The form of the equilibrium distribution function is given
in  (\ref{1.26a}). Then, it is
\begin{equation}\label{a1}
    \mathcal{P}_k^{(0)}(t^\prime)=\mathcal{P}_{e,k}(T^{(0)}) \, ,
\end{equation}
for $t^\prime\geq t_w$. Similarly to the decomposition carried out in  (\ref{1.15}), the equilibrium distribution corresponding to $T = T^{(0)}+ \Delta T$ can be written as
\begin{equation}\label{a4}
    \mathcal{P}_e (T)=\mathcal{P}_e (T^{(0)})+\mathcal{P}_e^{\prime} \, .
\end{equation}
It is
\begin{equation}\label{a5}
    \mathbb{W} \mathcal{P}_{e}(T)= \mathbb{W}^{(0)} \mathcal{P}_e^{\prime}+\mathbb{W}^{\prime} \mathcal{P}_e(T^{(0)})+\mathbb{W}^{\prime} \mathcal{P}_e^{\prime}=0 \, .
\end{equation}
In  the linear approximation being considered in this paper, the term $\mathbb{W}^{\prime} \mathcal{P}_e^{\prime}$
is neglected in the above relation, resulting that
\begin{equation}\label{a6}
    \mathbb{W}^{\prime}\mathcal{P}_e(T^{(0)})=- \mathbb{W}^{(0)} \mathcal{P}_e^{\prime} \, .
\end{equation}
use of this relationship into  (\ref{1.25}) yields
\begin{equation}\label{a7}
    R_{e,A}(t-t^\prime)=-\sum_i \sum_j \sum_k A_i T_{t-t^\prime}^{(0)}(i|j) W_{jk}^{(0)} \mathcal{P}_{e,k}^{\prime} \, .
\end{equation}
The transition probability $T_{t}^{(0)}(i|j)$ obeys the  ``backwards'' equation \cite{vk92}
\begin{equation}\label{a8}
    \frac{\partial}{\partial t}T_t^{(0)}(i|j)=\sum_k T_t^{(0)}(i|k) W_{kj}^{(0)} \, ,
\end{equation}
so that  (\ref{a7}) is equivalent to
\begin{eqnarray}
    R_{e,A}(t,t^\prime)&
    =-\sum_i\sum_j A_i \frac{\partial}{\partial t}\, T_{t-t^\prime}^{(0)}(i|j) \mathcal{P}_{e,j}^{\prime} \nonumber \\
    &= \sum_i\sum_j A_i \frac{\partial}{\partial t^\prime }\, T_{t-t^\prime}^{(0)}(i|j) \mathcal{P}_{e,j}^{\prime} \, . \label{a9}
\end{eqnarray}
From (\ref{1.26a}) and (\ref{a4}) it follows that
\begin{equation}\label{a11}
    \mathcal{P}_{e,j}^{\prime}= \mathcal{P}_{e,j}(T^{(0)}+\Delta T)-\mathcal{P}_{e,j}(T^{(0)})
    \approx -\left[ E_j-\langle E\rangle_e^{(0)} \right] \Delta\beta \mathcal{P}_{e,j}(T^{(0)}) \, ,
\end{equation}
where $\langle E\rangle_e^{(0)}$ is the average value of the energy in the equilibrium state at temperature $T^{(0)}$. By combining  (\ref{a9}) and (\ref{a11}) it is obtained that
\begin{equation}\label{a12}
    R_{e,A}(t-t^\prime)=-\Delta\beta \frac{\partial}{\partial t^\prime} \sum_i \sum_j A_i E_j T_{t-t^\prime}^{(0)}(i|j) \mathcal{P}_{e,j} (T^{(0)}) \, ,
\end{equation}
or,
\begin{equation}\label{a13}
    R_{e,A}(t-t^\prime)=-\Delta\beta \frac{\partial}{\partial t^\prime} \,\langle A(t) E(t^\prime) \rangle_e^{(0)}= \Delta\beta \frac{\partial}{\partial t}\ \langle A(t) E(t^\prime) \rangle_e^{(0)} \, ,
\end{equation}
where $\langle A(t) E(t^\prime) \rangle_e^{(0)}$ is the equilibrium time correlation function of the property $A$ and the energy of the system at temperature $T^{(0)}$. The general definition of these correlation functions is given in  (\ref{1.27a}). The above equation is a particular case of the well-known fluctuation-dissipation theorem, relating the response to a perturbation of a system at equilibrium with a time correlation function. A similar derivation of  (\ref{a13}) can be found in \cite{Be71}.

\section{Equilibrium time correlation functions}
\label{apb}

Suppose a master equation whose transition rates verify the detailed balance condition,
\begin{equation}\label{b1}
    W_{ij}\mathcal{P}_{e,j}=W_{ji}\mathcal{P}_{e,i} \, .
\end{equation}
where the equilibrium distribution has the canonical form given in  (\ref{1.26a}). Equilibrium
time correlation functions are defined by (see  (\ref{1.27a}))
\begin{equation}\label{b4}
    \langle A(t) B(t^\prime)\rangle_e=\langle A(t-t^\prime)B(0) \rangle_e=\sum_i \sum_j A_i B_j T_{t-t^\prime}(i|j) \mathcal{P}_{e,j} \, .
\end{equation}
 When detailed balance holds, the master equation can be solved by means of the eigenvalues and eigenvectors method \cite{vk92}. Let us briefly summarize it. Consider the eigenproblem
\begin{equation}\label{b5}
    \sum_j \mathbb{W}_{ij} \psi_j^{(\alpha)}=-\lambda^{(\alpha)} \psi_i^{(\alpha)} \, .
\end{equation}

For the sake of simplicity, it is assumed in the notation that  the eigenvalue $-\lambda^{(\alpha)}$ is non-degenerate, but the extension to the degenerate case is straightforward.
All the eigenvalues $-\lambda^{(\alpha)}$ are semi-defined negative, i.e. $\lambda^{(\alpha)} \geq 0$, and the only eigenvector corresponding to the null eigenvalue is the equilibrium distribution $\mathcal{P}_{e,i}$. Then it is \cite{vk92},
\begin{equation}\label{b10}
    T_{t}(i|j)=\sum_{\alpha} \frac{\psi_i^{(\alpha)}\psi_j^{(\alpha)}}{\mathcal{P}_{e,j}} e^{-\lambda^{(\alpha)}t}\, ,
\end{equation}
where the sum extends over all the eigenvectors of $\mathbb{W}$. Substitution of this expression into  (\ref{b4}) leads to
\begin{eqnarray}
    \langle A(t) B(t^\prime)\rangle_e&=\sum_{\alpha}\sum_i A_i \psi_i^{(\alpha)} \sum_j B_j \psi_j^{(\alpha)} e^{-\lambda^{(\alpha)}(t-t^\prime)} \nonumber \\
    &=\langle A\rangle_e \langle B\rangle_e +{\sum_{\alpha}}^{\prime} \Delta A^{(\alpha)} \Delta B^{(\alpha)} e^{-\lambda^{(\alpha)}(t-t^\prime)}\label{b11} \, .
\end{eqnarray}
Here the prime to the right of the summation symbol indicates that the equilibrium eigenvector is excluded, and
\begin{equation}\label{b12}
    \Delta A^{(\alpha)}=\sum_i A_i \psi_i^{(\alpha)} \, , \quad \Delta B^{(\alpha)}=\sum_i B_i \psi_i^{(\alpha)} \, .
\end{equation}
Putting $A=B$, the expression for the time autocorrelation function is obtained,
\begin{equation}\label{b13}
    \langle A(t)A(t^\prime)\rangle_e=\langle A\rangle_e^2 +{\sum_\alpha}^{\prime} \left( \Delta A^{(\alpha)}\right)^2 e^{-\lambda^{(\alpha)}(t-t^\prime)} \, ,
\end{equation}
and particularization for $t=t^\prime$ gives the second moment of the equilibrium fluctuations of the property $A$,
\begin{equation}\label{b14}
    \langle A^2\rangle_e = \langle A\rangle_e^2 +{\sum_\alpha}^{\prime} \left( \Delta A^{(\alpha)}\right)^2 \, .
\end{equation}
This equation shows that $\left( \Delta A^{(\alpha)}\right)^2$ is the contribution of the $\alpha$-th mode to the dispersion $\left( \Delta A \right)^2=\langle A^2\rangle_e-\langle A\rangle_e^2$. Equation (\ref{2.7}) follows by particularizing  (\ref{b13}) for $A$ being the energy of the system.

\end{document}